# Structure-Property-Performance Relationships of Cuprous Oxide Nanostructures for Dielectric Mie Resonance-Enhanced Photocatalysis

Ravi Teja A. Tirumala[⊥‡], Sunil Gyawali[§‡], Aaron Wheeler[⊥‡], Sundaram Bhardwaj Ramakrishnan[⊥], Rishmali Sooriyagoda[§], Farshid Mohammadparast[⊥], Susheng Tan[┼], A. Kaan Kalkan[⊥⊥], Alan D. Bristow[§]*, Marimuthu Andiappan*[,⊥]

Affiliations:

[⊥] School of Chemical Engineering, Oklahoma State University, Stillwater, OK, USA.

[§] Department of Physics and Astronomy, West Virginia University, Morgantown, WV, USA.

[⊥⊥] School of Mechanical and Aerospace Engineering, Oklahoma State University, Stillwater, OK, USA.

[┼] Department of Electrical and Computer Engineering and Petersen Institute of Nano Science and Engineering, University of Pittsburgh, Pittsburgh, PA, USA

[‡] These authors contributed equally

*Corresponding authors, Email: mari.andiappan@okstate.edu; alan.bristow@mail.wvu.edu

ABSTRACT

Nanostructured metal oxides, such as $Cu_2O$, $CeO_2$, $\alpha$-$Fe_2O_3$, and $TiO_2$ can efficiently mediate photocatalysis for solar-to-chemical energy conversion and pollution remediation. In this contribution, we report a novel approach, dielectric Mie resonance-enhanced photocatalysis, to enhance the catalytic activity of metal oxide photocatalysts. Specifically, we demonstrate that $Cu_2O$ nanostructures exhibiting dielectric Mie resonances can exhibit up to an order of magnitude higher photocatalytic rate as compared to $Cu_2O$ nanostructures not exhibiting dielectric Mie resonances. Our finite-difference time-domain (FDTD) simulation and experimental results predict a volcano-type relationship between the photocatalytic rate and the size of $Cu_2O$ nanospheres and nanocubes. Using transient absorption measurements, we reveal that a coherent electronic process associated with dielectric Mie resonance-mediated charge carrier generation is dominant in $Cu_2O$ nanostructures that exhibit higher photocatalytic rates. Although we experimentally demonstrate dielectric Mie resonance-enhanced photocatalysis using $Cu_2O$ particles here, based on our FDTD simulations, we anticipate the same can be achieved with other metal oxide photocatalysts, including $CeO_2$, $\alpha$-$Fe_2O_3$, and $TiO_2$.

**TOC GRAPHICS**

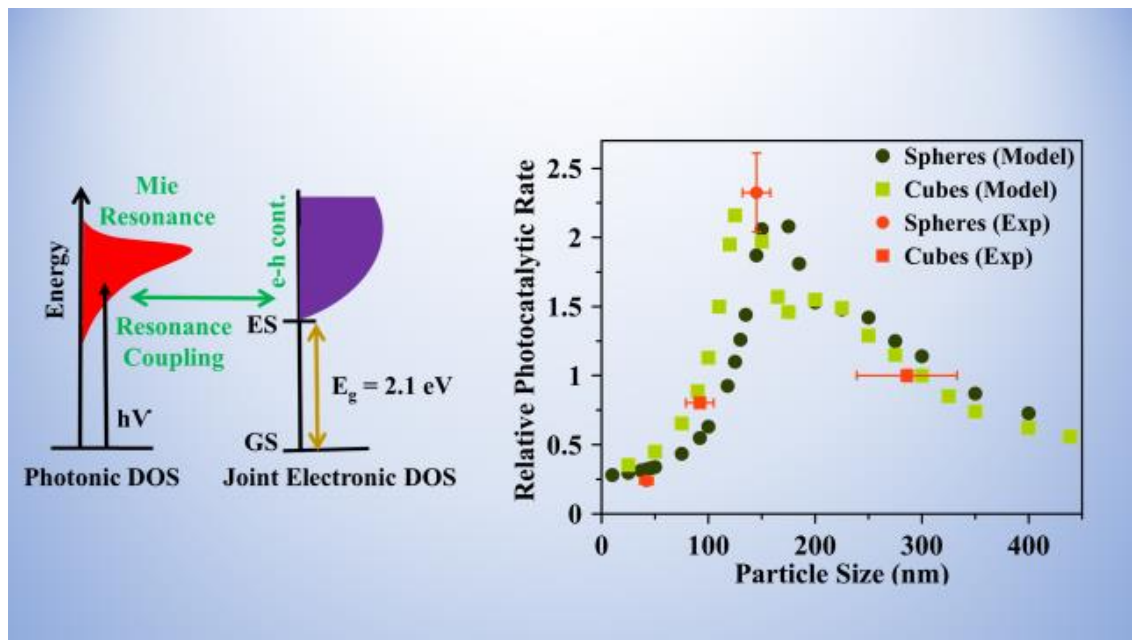

**MAIN TEXT**

**1. Introduction**

The discovery of photoelectrochemical water splitting using a $TiO_2$ photoelectrode by Fujishima and Honda in 1972, has opened the gateway for semiconductor-assisted photocatalysis.[1] In bandgap-facilitated photocatalysis using semiconductors, the photo-excited electrons and holes drive reduction and oxidation reactions, respectively (Figure 1a). For example, in photocatalytic water splitting, the photoexcited electrons reduce protons ($H^+$) to hydrogen ($H_2$) while photoexcited holes oxidize water to oxygen ($O_2$).

Researchers have explored the properties of semiconductors, such as cerium oxide ($CeO_2$), cuprous oxide ($Cu_2O$), hematite iron (III) oxide ($\alpha$-$Fe_2O_3$), and $TiO_2$ in the last four decades, for solar-to-chemical energy conversion and pollution mitigation applications.[2–7] However, some challenges still exist in using semiconductor-only-based photocatalysts for efficient conversion of solar energy to chemical energy. For instance, semiconductor nanostructures lack efficient harvesting of incoming photons because of their inherently poor absorption cross sections.[8] Plasmonic metal nanostructures (PMNs) have emerged as promising materials to overcome some of these limitations. Specifically, studies in the last decade have shown that hybrid and composite photocatalysts built on PMNs (e.g., Al, Ag, Au, and Cu) and metal oxide semiconductors (e.g., $Cu_2O$, $\alpha\text{-}Fe_2O_3$, and $TiO_2$) can exhibit enhanced photocatalytic activity as compared to semiconductor-only photocatalysts.[8–15]

The absorption cross sections of PMNs are four to five orders of magnitude higher than dye molecules.[16] This strong interaction of PMNs with the incident ultraviolet/visible (UV/Vis) light is due to the localized surface plasmon resonance (LSPR).[16–18] The PMN can therefore efficiently harvest the incident light and transfer the energy into the nearby metal oxide semiconductor via a number of electron- and energy-transfer pathways, including plasmon-induced hot electron transfer, nanoantenna effect, and plasmon-induced resonance energy transfer (Figure 1b).[8–11] The plasmonic resonance-mediated energy transfer from PMN into the nearby metal oxide semiconductor can result in an enhanced rate of generation of excited electrons and holes in the conduction and valence bands of the semiconductor, respectively (Figure 1b).[9] These enhanced rates of generation of charge carriers are shown to result in enhanced photocatalytic activity in PMN-metal oxide composite and hybrid photocatalysts.[8–12,19]

In our previous contribution,[20] we have reported that similar to the plasmonic resonances in PMNs, tunable dielectric resonances can be created in metal oxide particles by controlling their size and shape. Herein, we demonstrate that these dielectric resonances can be utilized to enhance the inherent photocatalytic activity of metal oxide photocatalysts (Figure 1c). In the plasmonic resonance-mediated photocatalytic approach shown in Figure 1b, two building blocks are required:

PMN to facilitate the plasmonic resonance and semiconductor for the bandgap-facilitated reduction and oxidation (redox) reactions. In contrast, in the proposed dielectric resonance-mediated photocatalytic approach shown in Figure 1c, a single metal oxide building block can serve the dual function and exhibit both the dielectric resonance behavior and the bandgap-facilitated redox reactions.

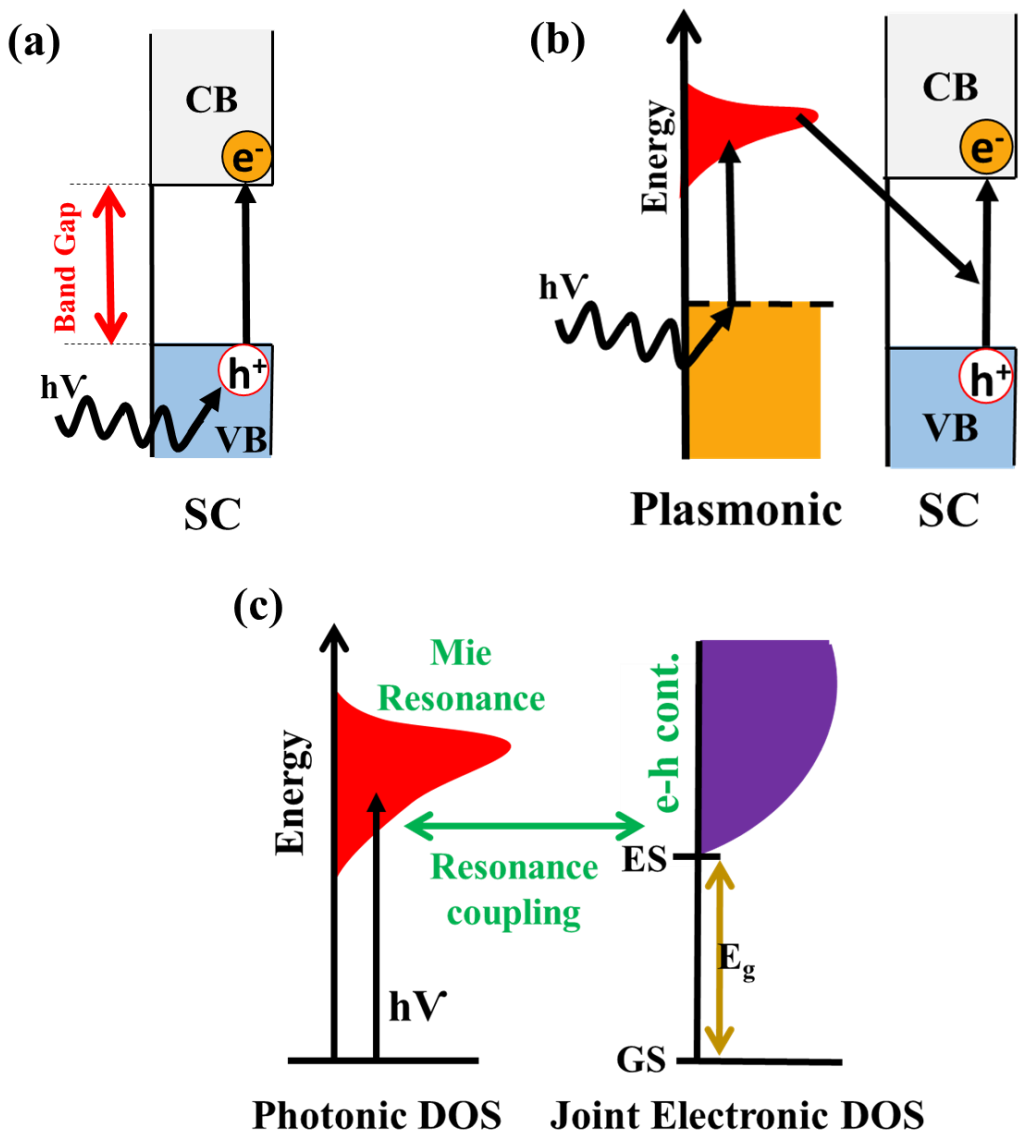

**Figure 1.** Schematic diagram illustrating (a) generation of a photoexcited electron-hole pair ($e^-$-$h^+$) in a semiconductor (SC), (b) plasmonic resonance-mediated generation of excited electrons and holes in a SC, and (c) the proposed dielectric resonance-mediated generation of excited electrons and holes in a SC. In (a), (b), and (c), excited electrons in the conduction band (CB) and excited holes in the valence band (VB) of SC can drive reduction and oxidation reactions.

The plasmonic and dielectric resonances are two distinct sub-categories of Mie resonances.[21,22] The plasmonic resonances occur for materials with negative values of the real part of the permittivity ($\varepsilon < 0$). The Mie resonances can also occur for materials with positive values of permittivity ($\varepsilon > 0$). When Mie resonances occur in the positive permittivity region, the resonances are known as dielectric resonances. Specifically, dielectric resonances can occur in materials with moderate (2.5-3.5) and high refractive index (>3.5).[23] Compared to plasmonic Mie resonances which exhibit only the electric multipole resonances (e.g., electric dipole, quadrupole. etc.), the dielectric Mie resonances can exhibit electric and magnetic multipole resonances upon light excitation.[20] Similar to the plasmonic resonances, the dielectric resonance wavelengths are also tunable with the particle size and shape.[20] At the resonance wavelengths, the dielectric nanostructures can act as optical nanoantenna and exhibit orders of magnitude enhancements for the electric and magnetic fields of the incident light. The nanoantenna and light trapping effects of dielectric nanostructures are utilized for dielectric resonance-enhanced light absorption, fluorescence, and Raman scattering.[21] For example, dielectric Mie resonances have been demonstrated for their use in controlling and enhancing light absorption for solar fuel generation and thin-film solar cell applications.[24,25]

Mie dielectric resonances exhibit themselves as absorption/extinction peaks in the UV-Vis-near infrared (UV-Vis-near IR) absorption/extinction spectra of medium- and high-refractive index dielectric particles.[20] These resonances, which are not present in their bulk counterparts, provide the opportunity for enhanced light harvesting in dielectric particles. For example, $FeS_2$ nanocubes with dielectric Mie absorption have been demonstrated to exhibit a higher photothermal conversion efficiency as compared to the small $FeS_2$ nanoparticles not exhibiting dielectric Mie absorption.[26] Herein, we demonstrate that $Cu_2O$ nanospheres and nanocubes, exhibiting dielectric Mie resonances, are associated with a higher photocatalytic rate as compared to smaller $Cu_2O$ nanoparticles not exhibiting dielectric Mie resonances.

$Cu_2O$ is a semiconductor with a dipole-forbidden direct gap of ~2.1 eV.[5–7,27] It is also a moderate refractive index material with the values of the real part of the refractive index in the range of ~ 2.6 - 3.1 (SI).[20] $Cu_2O$ is also known for exhibiting a relatively long lifetime of excited charge carriers (electrons and holes) that is on the order of milliseconds.[28,29] In our previous contribution[20], we have demonstrated that: (i) $Cu_2O$ nanocubes of edge lengths larger than ~100 nm exhibit strong dielectric resonances in the Vis-near IR regions, (ii) the extinction cross section of these $Cu_2O$ cubic particles are comparable to or larger than those of plasmonic Ag particles of similar sizes, and (iii) smaller $Cu_2O$ spherical and cubic nanoparticles of sizes less than 100 nm exhibit light absorption features similar to their bulk counterparts and does not exhibit any Mie resonances in the Vis-near IR regions. In this contribution, we demonstrate that larger $Cu_2O$ quasi-spherical nanoparticles of 145 nm average diameter exhibit higher visible-light photocatalytic activity for methylene blue (MB) dye degradation as compared to smaller $Cu_2O$ quasi-spherical nanoparticles of 42 nm average diameter. In the remainder of this report, we will refer to these three samples as '145-nm spheres' and '42-nm spheres' for the sake of brevity, although the particle shapes are not exact cubes and spheres, respectively. We attribute the higher photocatalytic rate of 145-nm spheres to their dielectric Mie resonances, which are not present in the 42-nm spheres. The results from the photoreactor studies are supported by the transient-absorption measurements that are used to identify the incoherent or coherent charge carrier dynamics that are markedly different for the smaller and larger spheres. We have also provided structure-photocatalytic performance relationships that capture the proposed dielectric Mie resonance-enhanced photocatalysis in $Cu_2O$ nanospheres and nanocubes. Specifically, our finite-difference time-domain (FDTD) simulation and experimental results predict a volcano-type relationship between the relative photocatalytic rate and the size of $Cu_2O$ nanospheres and nanocubes.

## 2. Experimental Section

Syntheses and characterizations of $Cu_2O$ nanospheres and $Cu_2O$ nanocubes: Microemulsion technique is used for the synthesis of smaller $Cu_2O$ quasi-spherical nanoparticles of average diameters in the range of 35-45 nm.[30,31] For the syntheses of larger $Cu_2O$ quasi-spherical particles of 145 nm average diameter and $Cu_2O$ nanocubes of tunable edge lengths in the range of 90-350 nm, chemical reduction method is used.[20] The synthesis temperatures of ~55 ˚C and ~20 ˚C are used for the synthesis of larger $Cu_2O$ quasi-spherical nanoparticles and nanocubes, respectively. The photocatalytic activity of $Cu_2O$ nanospheres and nanocubes of different sizes towards methylene blue (MB) degradation are investigated in the solution phase using dimethylformamide (DMF) as a solvent. The detailed procedures followed for the syntheses and characterizations of $Cu_2O$ nanoparticles, and FDTD simulations, photocatalytic experiments, and transient absorption measurements are provided in the supporting information (SI).

## 3. Results

Figure 2a-d shows the representative TEM image of 42-nm $Cu_2O$ nanospheres and SEM images of 145-nm spheres, 92-nm nanocubes, and 286-nm nanocubes, respectively. The $Cu_2O$ phase of these cubical and spherical particles is confirmed by XRD analysis (e.g., Figure S1a-b in SI). Their photocatalytic performances toward MB degradation are evaluated under visible light irradiation. In these studies, the photocatalytic MB dye degradations are carried out for the same weight load of $Cu_2O$ nanoparticles. Green LED lamps that emit light with a peak wavelength at 519 nm are used as the visible light source. A representative picture of the photoreactor is provided in Figure S2 in SI.

In Figure 3a, we provide the spectrum of green LEDs and the absorption spectrum of MB. As seen from Figure 3a, there is minimal overlap between the spectrum of the light source (i.e., green LEDs) and the absorption spectrum of methylene blue. This minimal overlap will ensure that the MB dye degradation due to direct dye excitation is minimal. In Figure 3b, we show the extent of MB dye degradation (i.e., the ratio of the concentration of MB to its initial concentration, $C/C_0$) by 42-nm and 145-nm $Cu_2O$ nanospheres and 92-nm and 286-nm $Cu_2O$ nanocubes as a function of visible light irradiation time. As seen from Figure 3b, the 145-nm $Cu_2O$ nanospheres exhibit the highest MB degradation rate while 42-nm $Cu_2O$ nanospheres exhibit the lowest rate among the $Cu_2O$ nanostructures investigated in the photocatalytic experiments. Hence, despite their enhanced photocatalytic rate, the 145-nm spheres have only 0.29 times of the total surface area of the 42-nm spheres. We also performed photocatalytic experiments under visible light irradiation in the absence of $Cu_2O$ particles and found no significant MB degradation under these blank conditions.

To confirm that the MB degradations shown in Figure 3b are not due to light-induced heating, we measured the change in temperature of the reaction mixture as a function of irradiation time. We found that visible light irradiation increases the temperature of the reaction mixture from ~20 °C to a maximum of ~28 °C (Figure S3a-e in SI). Specifically, we observed a similar extent of increase in temperature for the photocatalytic experiments performed using the 42-nm and 145-nm $Cu_2O$ nanospheres, 92-nm and 286-nm cubes, and under blank conditions in the absence of $Cu_2O$ nanoparticles (Figure S3a-e in SI). The heat localization in our system is also expected to be minimal since the photocatalytic experiments are performed in the liquid medium under stirring.[32] We also carried out heating experiments with $Cu_2O$ nanospheres and nanocubes under dark conditions. In these heating experiments, the temperature of the reaction mixture is maintained at 60 °C. For all tested experimental conditions, the heating trials revealed no significant MB degradation in the absence of light exposure across an extended period (e.g., Figure S3f in SI).

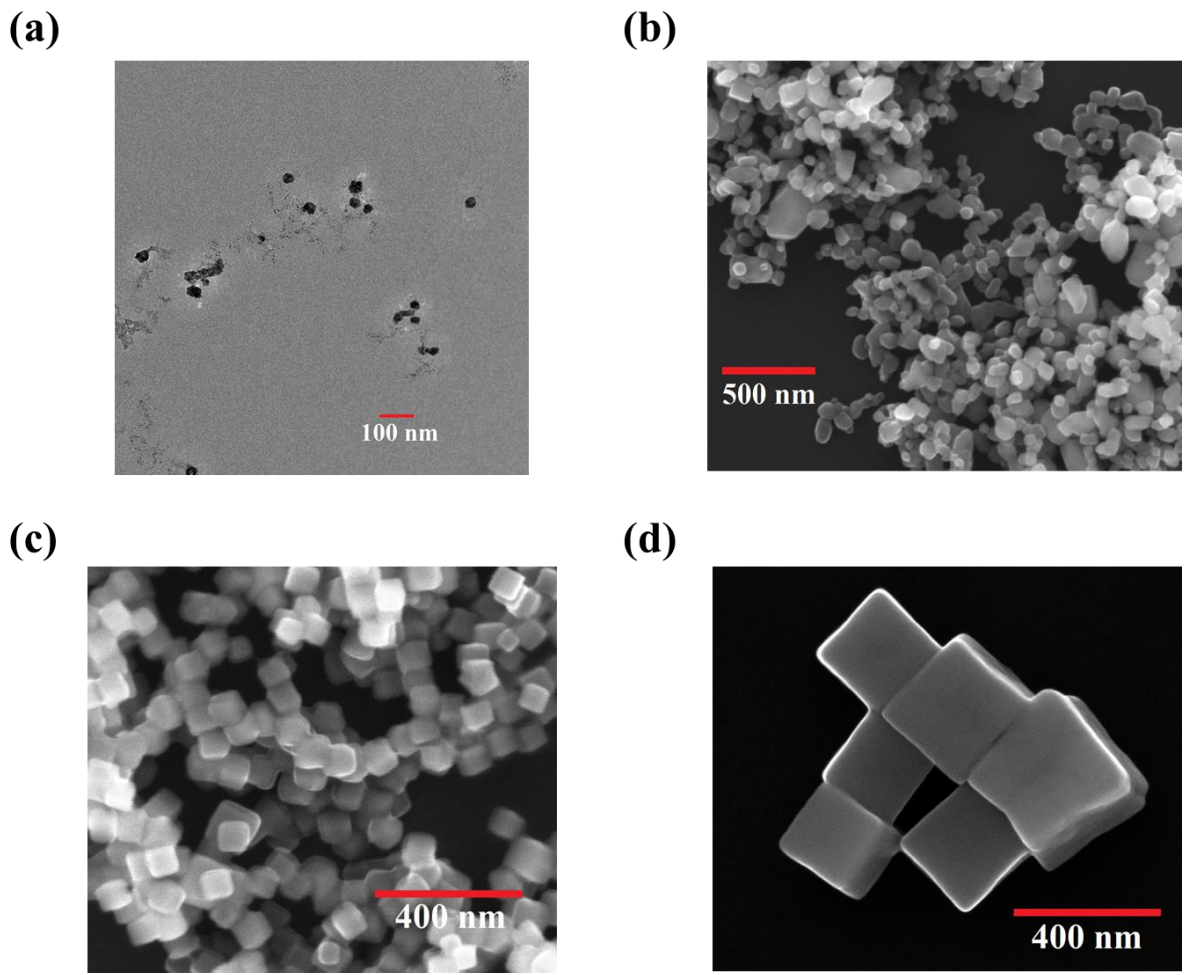

**Figure 2. (a)** Representative TEM image of smaller Cu$_2$O quasi-spherical nanoparticles of 42 ± 6 nm diameter. **(b-d)** Representative SEM images of **(b)** larger Cu$_2$O quasi-spherical nanoparticles of 145 ± 13 nm diameter, **(c)** Cu$_2$O nanocubes of 92 ± 13 nm edge length, and **(d)** Cu$_2$O nanocubes of 286 ± 47 nm edge length.

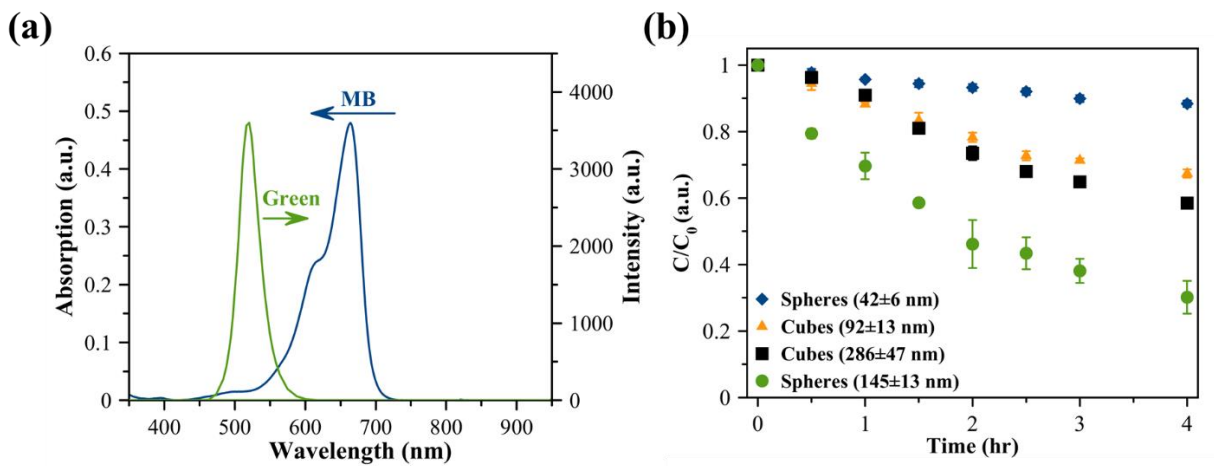

**Figure 3. (a)** Absorption spectrum of MB (left ordinate) and spectrum of green LED light source used for photocatalytic degradation of MB (right ordinate). **(b)** Plot of $C/C_0$ versus irradiation time for photocatalytic degradation of MB using 42-nm $Cu_2O$ nanospheres (blue diamonds), 92-nm $Cu_2O$ nanocubes (orange triangles), 286-nm $Cu_2O$ nanocubes (black squares), and 145-nm $Cu_2O$ nanospheres (green circles).

To show that the difference in the photocatalytic rates of $Cu_2O$ nanostructures in Figure 3b are due to their dielectric Mie resonances, in Figure 4a-b, we show the UV-Vis extinction spectra of 42-nm and 145-nm $Cu_2O$ nanospheres and 92-nm and 286-nm $Cu_2O$ nanocubes, respectively. As seen from Figure 4a, the 42-nm spheres, being consistent with what is expected for their bulk counterparts, exhibit no resonance peaks in the UV-Vis-near IR extinction spectra. In contrast, the 145-nm nanospheres in Figure 4a, and 92-nm and 286-nm $Cu_2O$ nanocubes in Figure 4b exhibit Mie resonance peaks in the visible to near IR regions. To understand the nature of these resonance peaks, we performed FDTD simulations. The representative simulated extinction spectra of the 42-nm $Cu_2O$ nanosphere and 286 nm-nanocube are shown in Figure S4a in SI. 4c-d. The features observed in the simulated extinction spectra of the 42-nm spheres and 286-nm cubes are consistent with the experimentally measured extinction spectra of the respective $Cu_2O$ nanoparticles shown in Figure 4a-b. The magnetic and electric field distributions at different wavelengths across the resonance peaks of the 286-nm nanocubes are also provided in Figure S4b-c in SI. The field distributions confirm that the resonance peaks observed in the extinction spectra of the 286-nm cube are due to the dielectric Mie resonances.[20] From the field distribution maps, we assign the lowest energy resonance peak observed in the extinction spectrum of the $Cu_2O$ cube to the combination of electric and magnetic dipoles (Figure S4b-c in SI).[20] Similarly, we assign the second lowest energy resonance peak and higher-order resonance peaks to the combination of electric and magnetic quadrupoles and the combination of higher-order electric and magnetic resonance modes, respectively.[20] From the Figure 3b and Figure 4a-b, it is clear that 145-nm $Cu_2O$ nanospheres and 92-nm and 286 nm nanocubes, exhibiting dielectric Mie resonances, are associated with a higher photocatalytic rate as compared to 42-nm $Cu_2O$ nanospheres not exhibiting dielectric Mie resonances.

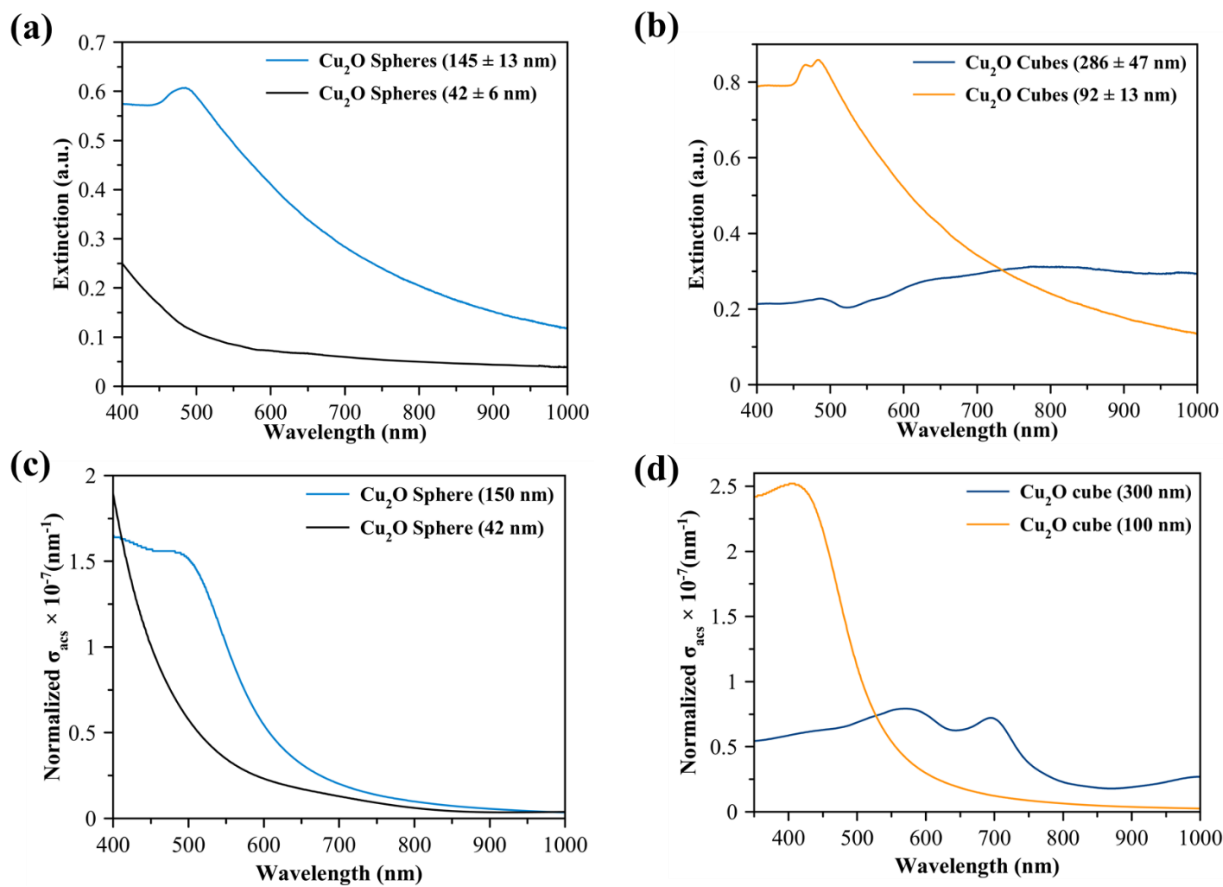

**Figure 4. (a-b)** Experimentally measured UV-Vis extinction spectra of **(a)** $Cu_2O$ quasi-spherical nanoparticles of 42 ± 6 nm and 145 ± 13 nm diameters, and **(b)** $Cu_2O$ nanocubes of 92 ± 13 nm and 286 ± 47 nm edge lengths. **(c-d)** FDTD-simulated volume-normalized absorption cross section ($\sigma_{acs}$) as a function of incident light wavelength for **(c)** $Cu_2O$ nanospheres of 42 and 150 nm diameters, and **(d)** $Cu_2O$ nanocubes of 100 nm and 300 nm edge lengths.

To develop the structure-property-photocatalytic performance relationships of $Cu_2O$ nanostructures, we propose a rational approach to predict the photocatalytic reaction rate on $Cu_2O$ nanospheres and nanocubes of different sizes. In this approach, it is approximated that the photocatalytic rate on $Cu_2O$ nanostructures is proportional to their charge carrier generation capacity. From this approximation, the photocatalytic rate at a given incident light wavelength is proportional to the incident light intensity and the volume-normalized absorption cross section of the $Cu_2O$ nanostructures. The FDTD simulations are used to predict this key optical property, i.e., volume-normalized absorption cross section of the $Cu_2O$ nanostructures of different sizes as a function of incident light wavelength. The photocatalytic MB dye degradation studies in our system are carried out for the same weight load of $Cu_2O$ nanospheres and nanocubes. Therefore, the comparison of the volume-normalized absorption cross section values of $Cu_2O$ nanostructures is a good descriptor to predict their relative light absorption capacity in our system (see SI for more details). In Figure 4c-d, we show the representative simulated absorption spectra of the 42-nm and 150 nm-nanospheres and 100-nm and 300-nm nanocubes, respectively. The simulated absorption spectra of a wide range of sizes are also provided in Figure S5a-b in SI. The y-axis values in these figures correspond to the volume-normalized absorption cross section. The overlap between the FDTD-simulated absorption spectra of the $Cu_2O$ nanostructures and the incident light spectrum is used as a descriptor to predict the photocatalytic rate of the respective nanostructures. We used this approach to predict the relationship between the size of the $Cu_2O$ nanospheres and nanocubes and their photocatalytic performance for the photocatalytic degradation of methylene blue molecules. The representative results are shown in Figure 6. The photocatalytic rate on 286-nm $Cu_2O$ nanocubes is used as a reference to calculate the relative rate of $Cu_2O$ nanospheres and nanocubes of different sizes in Figure 5.

The experimentally measured kinetic data shown in Figure 3b are fitted to obtain the apparent first-order rate constant values (see Figure S5c-f and Table S1 in SI). The experimental value for the relative photocatalytic rate of $Cu_2O$ nanospheres and nanocubes of a given size is obtained from the ratio of the rate constant value of the respective nanostructures to that of 286-nm $Cu_2O$ nanocubes. As seen from Figure 6, our simulation and experimental results predict a volcano-type relationship between the photocatalytic rate and the size of $Cu_2O$ nanospheres and nanocubes. Specifically, among the $Cu_2O$ nanospheres and nanocubes investigated in our photocatalytic experiments, the absorption spectrum of the 145-nm nanosphere provides a better overlap with the green LED spectrum in the ~490-560 nm wavelength region due to the dielectric Mie resonance peak. This better overlap results in the highest photocatalytic rate by 145-nm nanospheres. The dielectric Mie resonance-enhanced absorption causes the 145-nm nanospheres to exhibit an order of magnitude (9.76 times) higher photocatalytic rate as compared to 42-nm nanospheres, in which the Mie resonance is absent (see Figure 5 and Table S1 in SI).

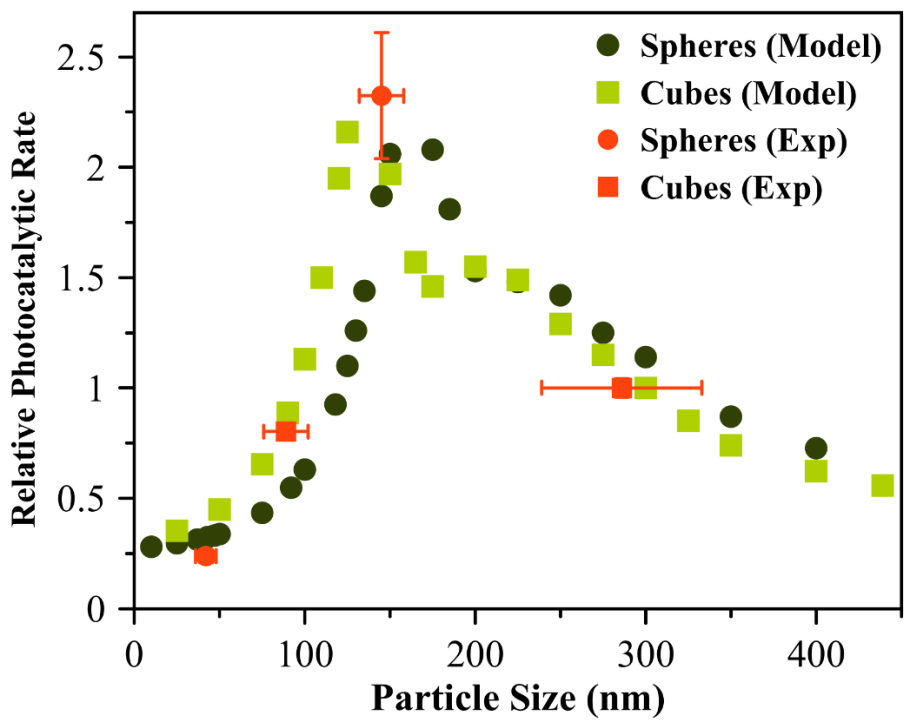

**Figure 5.** Volcano plot showing the predicted and experimentally measured relative photocatalytic rate as a function of size of $Cu_2O$ nanospheres and nanocubes. The photocatalytic rate on $Cu_2O$ nanocubes of 286 nm average edge length is used as a reference.

To further illustrate the Mie resonance-mediated photocatalysis, we show the transient absorption spectra of $Cu_2O$ nanocubes and $Cu_2O$ nanospheres in Figure 6a-b. Specifically, Figure 6a shows the transients for short time periods for two sets of $Cu_2O$ nanospheres of 37 ± 6 nm and 43 ± 5 nm diameters and two sets of $Cu_2O$ nanocubes of 297 ± 38 nm and 292 ± 39 nm edge lengths. Using the pump and probe conditions described in the methods section, the excitation in either sample is sufficiently above any absorption edge of the $Cu_2O$, creating excited-state charge carriers. The probe pulse then senses the excited-state carriers through free-carrier absorption of the pumped carriers, pushing them further into their respective bands. For the positive delay time both $Cu_2O$ nanospheres and $Cu_2O$ nanocubes exhibit a three-component exponential decay with typical delay times corresponding to a few picoseconds, a few tens of picoseconds, and a few hundred picoseconds. Fits are overlaid on the transient-absorption data shown for the full-data range for one of each sample type in Figure 6b and extracted parameters are given in Table 1.

Markedly, the negative delay time response from the 292-nm and 297-nm $Cu_2O$ nanocubes and 145-nm $Cu_2O$ nanospheres does not have the fast rise that is seen exhibited by the 37-nm and 43-nm $Cu_2O$ nanospheres. In the latter case, the rise is limited to the autocorrelation of the pump and probe pulses in the sample, typically the same order of magnitude as the laser pulse width (~100 fs). The response of 292-nm and 297-nm $Cu_2O$ nanocubes and 145-nm $Cu_2O$ nanospheres is indicative of a coherent response, like that seen as perturbed free induction in bleachable dyes,[33] GaAs quantum well,[34] single interfacial quantum dots,[35] and $Au@SiO_2@Cu_2O$ core-shell nanoparticles that demonstrate plasmon-induced resonance energy transfer (PIRET, Figure 1b).[9] Experimentally, these processes are observed because the weaker probe pulse imparts its coherent wavefront into the samples, which evolves over time, but has sufficient "*memory*" to then scatter some of the stronger pump light into the probe-detection direction, hence making this coherent process visible during negative delay times. The negative delay time response is fit with a double exponential decay, which reveals a faster component of a similar rise time as that of the nanospheres, which do not exhibit the protracted rise and a slower (~5 ps) response that is associated with the coherent energy transfer mechanism. Extracted fit parameters are also shown in Table 1.

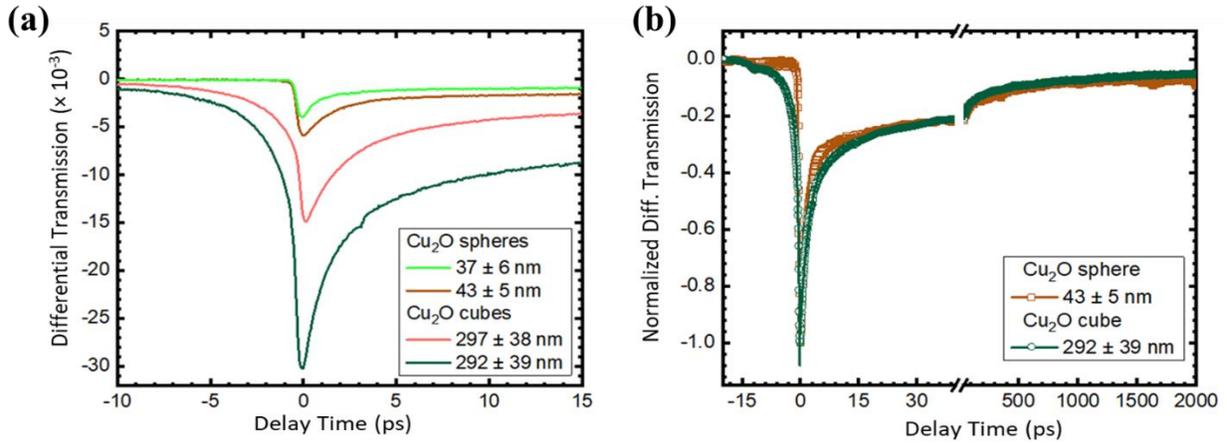

**Figure 6.** (a) Temporal response of the differential transmission of $Cu_2O$ spheres of $37 \pm 6$ nm and $43 \pm 5$ nm diameters and $Cu_2O$ cubes of $297 \pm 38$ nm and $292 \pm 39$ nm edge lengths. (b) Temporal response of the normalized differential transmission with exponential fits overlaid for $Cu_2O$ spheres of $43 \pm 5$ nm diameter and $Cu_2O$ cubes of $292 \pm 39$ nm edge length. A pump pulse of 400 nm wavelength and a probe pulse of 800 nm wavelength are used in (a) and (b).

**Table 1**. Extracted fitting parameters for the differential transmission data.

| $Cu_2O$ Sample | $t_{r1}$ (ps) | $t_{r2}$ (ps) | $t_{d1}$ (ps) | $t_{d2}$ (ps) | $t_{d3}$ (ps) |
|---|---|---|---|---|---|
| Spheres ($37 \pm 6$ nm) | 0.41 | - | 0.36 | 34.0 | 2282.9 |
| Spheres ($43 \pm 5$ nm) | 0.64 | - | 0.95 | 40.0 | 2073.6 |
| Spheres ($145 \pm 41$ nm) | 0.64 | 4.8 | 2.10 | 20.33 | 2800.0 |
| Cubes ($292 \pm 39$ nm) | 0.72 | 5.4 | 2.37 | 37.1 | 669.6 |
| Cubes (297 ± 38 nm) | 0.53 | 4.5 | 2.43 | 37.5 | 672.0 |

The coherent signature observed in the transient absorption of the 292-nm and 297-nm $Cu_2O$ nanocubes and 145-nm $Cu_2O$ nanospheres is expected because the dielectric Mie resonance-mediated charge carriers generation in dielectric $Cu_2O$ nanoparticles (Figure 1c) is most likely a coherent process, in a similar way to the PIRET-mediated charge carriers generation in the $Cu_2O$ shell of Au@$SiO_2$@$Cu_2O$ core-shell nanoparticles (Figure 1b).[9] Hence, the results of transient absorption measurements shown in Figure 6a-b and Table 1 provide direct evidence of a coherent process in the 292-nm and 297-nm $Cu_2O$ nanocubes and 145-nm $Cu_2O$ nanospheres (i.e., which also exhibit an enhanced photocatalytic activity).

To illustrate the photocatalytic mechanism that is responsible for the MB degradation on $Cu_2O$ cubes, we investigated the possible role of the solvent and superoxide ($O_2^-$) on MB degradation. To investigate the solvent dependency on MB degradation, we performed photocatalytic experiments using two different solvents, DMF and ethanol, and the representative results for 286-nm $Cu_2O$ nanocubes are shown in Figure 7a. In the solvent-dependent studies, $Cu_2O$ nanocubes exhibited a similar or slightly faster MB degradation in DMF compared to ethanol. We attribute the observed slightly faster MB degradation to the expected higher solubility of dissolved oxygen in DMF compared to ethanol. Previous studies have demonstrated that MB degradation on semiconductor photocatalysts can occur via a superoxide-mediated mineralization mechanism.[36,37]

To investigate whether photocatalytic MB degradation occurs via a superoxide-mediated mechanism, we performed the photocatalytic experiments with and without benzoquinone, a well-known scavenger of the superoxide.[38] The results from these experiments are shown in Figure 7b. The direct evidence of the superoxide-mediated mechanism can be gathered from Figure 7b, with the benzoquinone successfully inhibiting the MB degradation via scavenging of the superoxide. Based on these findings, we propose the following photocatalytic mechanism for the MB degradation on large $Cu_2O$ cubes. This mechanism, schematically illustrated in Figure 7c, involves the excitation of dielectric Mie resonances by the incident photons. The energy stored in Mie extinction dissipates into Mie scattering and Mie absorption. A major fraction of the energy that corresponds to Mie absorption transfers into and results in the coherent generation of excited electrons ($e^-$) and holes ($h^+$) in the conduction and valence bands of $Cu_2O$, respectively. The excited electrons from the conduction band reduce the dissolved oxygen ($O_2$) into superoxide ($O_2^-$). The superoxide then reacts with the MB molecule and forms several intermediate products, one of them being a carboxylic acid ($RCOO^-$) intermediate.[36,37] This intermediate can be oxidized by the excited hole ($h^+$) in the valence band into degradation products.[36,37]

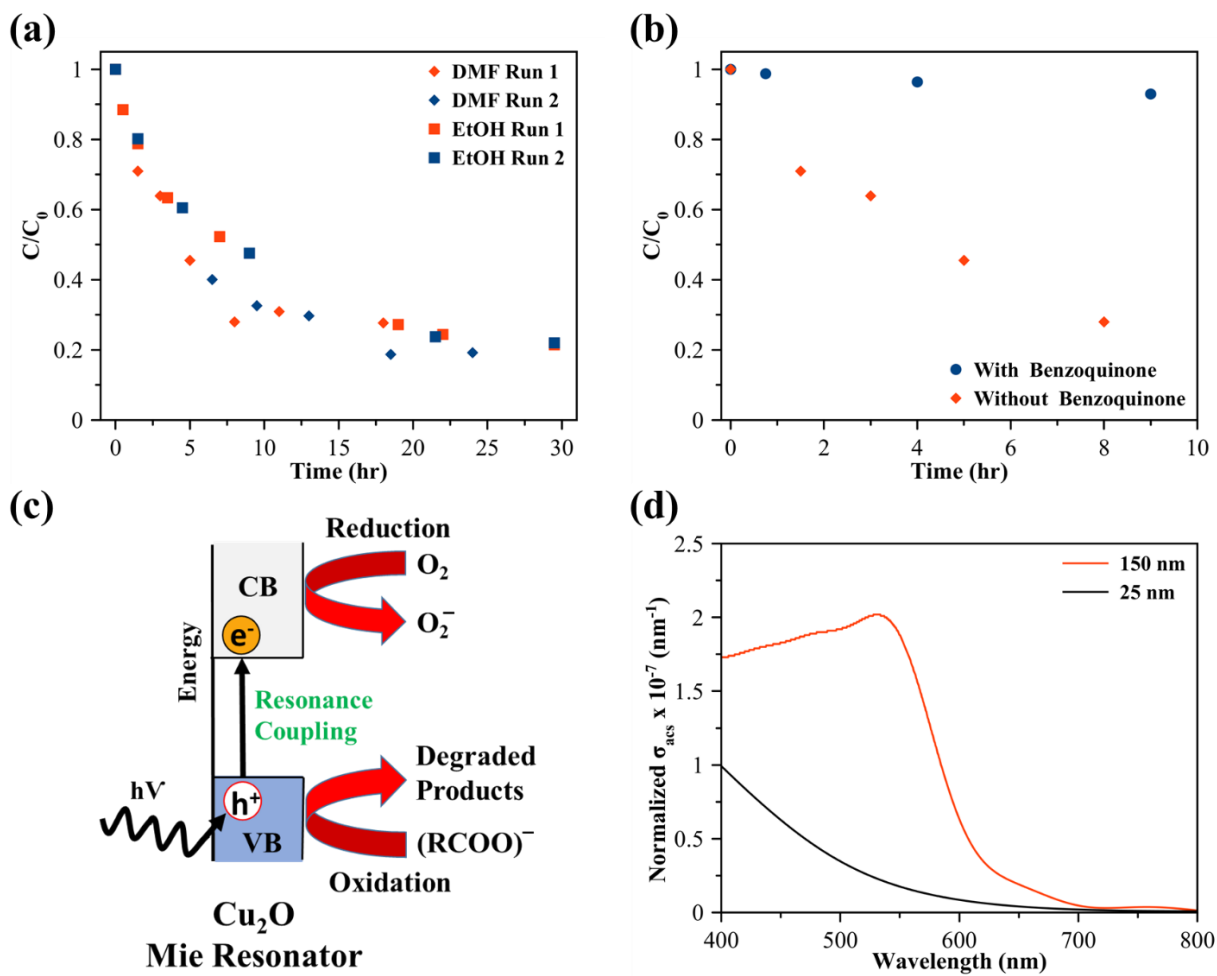

**Figure 7. (a)** Plot of $C/C_0$ versus irradiation time for photocatalytic degradation of MB in DMF (red and blue diamonds) and ethanol (red and blue squares) using 286-nm $Cu_2O$ nanocubes. **(b)** Plot of $C/C_0$ versus irradiation time for photocatalytic degradation of MB in the presence (blue circles) and absence (red diamonds) of benzoquinone using 286-nm $Cu_2O$ nanocubes in DMF. **(c)** Schematic diagram illustrating the proposed dielectric resonance-enhanced photocatalytic degradation of MB that occurs via a superoxide ($O_2^-$)-mediated mechanism. **(d)** FDTD-simulated volume-normalized absorption cross section ($\sigma_{acs}$) as a function of incident light wavelength for a larger $\alpha$-$Fe_2O_3$ nanosphere of 150 nm diameter and a smaller $\alpha$-$Fe_2O_3$ nanosphere of 25 nm diameter.

The dielectric resonance-enhanced photocatalysis demonstrated herein can be potentially applied to other metal oxide photocatalysts. For example, $CeO_2$, $\alpha$-$Fe_2O_3$, and $TiO_2$ semiconductors are also moderate and high refractive index materials with the values of the real part of the refractive index in the range of 2.1-2.5, 2.7-3.3, and 1.4-5.4, respectively (SI). To investigate whether these metal oxide particles exhibit dielectric resonance characteristics similar to that of $Cu_2O$, we performed FDTD simulations to simulate the absorption spectra of different sizes of $CeO_2$, $\alpha$-$Fe_2O_3$, and $TiO_2$ particles. The representative spectra are shown in Figure 7d and also in Figure S7a-c in SI. Our FDTD simulations show that $CeO_2$, $\alpha$-$Fe_2O_3$, and $TiO_2$ particles exhibit dielectric resonance characteristics that are similar to $Cu_2O$ particles. Specifically, our findings show that small nanoparticles such as spherical nanoparticles of 25 nm diameter exhibit light absorption features similar to their bulk counterparts and do not exhibit any Mie resonances in the near UV-Vis-near IR regions. In contrast, large nanoparticles such as spherical nanoparticles of 150 nm diameters exhibit strong dielectric resonances. For example, Figure 7d shows the volume-normalized absorption cross section values as a function of incident light wavelengths for 25 and 150 nm spherical $\alpha$-$Fe_2O_3$ particles. As seen from Figure 7d, in the visible region, 150 nm spherical nanoparticle exhibits relatively higher light absorption capacity than small spheres of 25 nm diameter. Similar to the dielectric resonance-enhanced photocatalysis demonstrated in larger $Cu_2O$ cubic particles in this contribution, the dielectric resonance features of larger $CeO_2$, $\alpha$-$Fe_2O_3$, and $TiO_2$ particles can be possibly explored to enhance their inherent photocatalytic activity.

## 4. Conclusions

The present work demonstrates dielectric resonance-enhanced photocatalysis in $Cu_2O$ cubical and spherical particles. Specifically, 145-nm $Cu_2O$ spherical nanoparticles with dielectric resonances exhibit a higher photocatalytic rate for MB degradation in comparison to 42-nm $Cu_2O$ spherical nanoparticles not exhibiting dielectric resonances. These results are supported by the transient absorption measurements that differentiate the charge carrier dynamics of 145-nm and 42-nm $Cu_2O$ nanospheres, indicating an optical resonance process that involves the coherent generation of charge carriers in the 145-nm nanospheres. This present work also presents volcano-type structure-performance relationships between the size of $Cu_2O$ nanostructures and their relative photocatalytic rates. The dielectric resonance-enhanced photocatalytic approach demonstrated in this study using $Cu_2O$ nanoparticles is expected to translate to other metal oxide photocatalysts such as $CeO_2$, $\alpha$-$Fe_2O_3$, and $TiO_2$ that exhibit size- and shape-dependent dielectric resonances. The dielectric Mie resonance-mediated charge carrier generation in the metal oxide semiconductors has a number of advantages over plasmonic resonance-mediated charge carrier generation. Specifically, in the latter case, the plasmonic Mie mode decays to a hot electron (and hole). In comparison to the lifetime of hot electron (and hole), the electron-hole pair lifetime in the metal oxide semiconductor is longer, which will translate to a much more efficient photocatalysis. Therefore, the dielectric resonance-enhanced photocatalysis demonstrated in our study opens a new avenue for solar light harvesting and photocatalysis.

ASSOCIATED CONTENT

**Supporting Information**

Detailed descriptions of photocatalysts syntheses and characterization procedures, photocatalytic experimental procedures, transient absorption measurements, FDTD simulations and supporting figures are listed and supplied in Supporting Information.

AUTHOR INFORMATION

**Corresponding authors, Email:** mari.andiappan@okstate.edu; alan.bristow@mail.wvu.edu**Corresponding authors, Email:** mari.andiappan@okstate.edu; alan.bristow@mail.wvu.edu

# Structure-Property-Performance Relationships of Cuprous Oxide Nanostructures for Dielectric Mie Resonance-Enhanced Photocatalysis

*Ravi Teja A. Tirumala$^{\perp\ddagger}$, Sunil Gyawali$^{\S\ddagger}$, Aaron Wheeler$^{\perp\ddagger}$, Sundaram Bhardwaj Ramakrishnan$^{\perp}$, Rishmali Sooriyagoda$^{\S}$, Farshid Mohammadparast$^{\perp}$, Susheng Tan$^{\dagger}$, A. Kaan Kalkan $^{\parallel}$, Alan D. Bristow$^{\S}$*, Marimuthu Andiappan*$^{,\perp}$*

Affiliations:

$^{\perp}$ School of Chemical Engineering, Oklahoma State University, Stillwater, OK, USA.

§ Department of Physics and Astronomy, West Virginia University, Morgantown, WV, USA.

$^{\parallel}$ School of Mechanical and Aerospace Engineering, Oklahoma State University, Stillwater, OK, USA.

╪ Department of Electrical and Computer Engineering and Petersen Institute of NanoScience and Engineering, University of Pittsburgh, Pittsburgh, PA, USA.

‡ These authors contributed equally

*Corresponding authors, Email: mari.andiappan@okstate.edu; alan.bristow@mail.wvu.edu

# I. Syntheses and characterizations of Cu₂O nanospheres and nanocubes

**Syntheses of Cu₂O quasi-spherical nanoparticles**

Smaller Cu₂O nanoparticles of quasi-spherical shape and diameters in the range of 35-45 nm were synthesized using the microemulsion technique at room temperature (~20 ˚C). In this synthesis method, 54.5 mL of n-heptane (oil phase) and polyethylene glycol-dodecyl ether (Brij, average Mn ~362) as surfactant are added to a 250 mL round bottom flask and allowed to stir at 550 rpm. 5.4 mL of 0.1 M copper nitrate aqueous solution is added to this mixture and 1 M hydrazine solution (5.4 mL) is added as a reducing agent. This oil in water micro-emulsion utilizes the fact that small sizes of reverse micelles are formed in which the Cu₂O nanoparticles are formed. The mixture is allowed to stir for 12 hours after which acetone is added to break the emulsion and centrifuged. These nanoparticles are washed three times (sonicated and centrifuged) to remove the surfactant and to obtain Cu₂O spherical nanoparticles that are ready to be used as photocatalyst for methylene blue dye degradation.

The chemical reduction method at a synthesis temperature of 55 ˚C was used for the synthesis of larger Cu₂O quasi-spherical nanoparticles of 145 ± 13 nm diameter. 50 mL of 10 mM CuCl₂ aqueous solution was prepared in a 100 mL round bottom flask. The mixture was allowed to stir at 900 rpm at 55°C. 5 mL of 2 M NaOH solution was added to the mixture and allowed to stir under constant heating (55 ˚C) for 30 minutes, followed by the addition of 5 mL of 0.6 M ascorbic acid aqueous solution as reducing agent. The synthesis mixture was allowed to stir for 5 hours. The resulting nanoparticles were separated by washing them in DI water and ethanol three times each to remove all residue from the synthesis mixture. The washed and clean larger Cu₂O spherical nanoparticles were then used in the photocatalytic experiments.

**Syntheses of Cu₂O nanocubes**

Larger Cu₂O nanocubes with average edge lengths in the range of 280-300 nm were synthesized using a chemical reduction method performed at room temperature (~20 ˚C). In this method, we first prepared a copper source of 30 mL of 0.0032 M aqueous CuCl₂ solution. This solution is put into a three-neck round bottom flask, which is put in an inert environment filled with nitrogen. We added 1 mL of 0.35 M aqueous NaOH solution to this solution at room temperature, which should result in the creation of blue-colored Cu(OH)₂ colloids almost immediately. The sodium ascorbate (reducing agent) was then added in 1 mL increments. The solution subsequently became orangish-yellow, suggesting that Cu₂O cubic particles are formed. The synthesis duration was a period of one hour after which, the Cu₂O cubes were washed using ethanol three times (sonicated and centrifuged). The as-prepared Cu₂O nanoparticles are suspended in 4 mL of the reaction solvent as per the reaction procedure mentioned in section II and are used in photocatalytic experiments.

To synthesize medium-sized Cu₂O nanocubes with average edge lengths in the range of 90-120 nm, chemical reduction method was used at room temperature (~20 ˚C). A 500 mL-three-neck round bottom flask (reactor) at room temperature (~20 ˚C) is flushed with N₂ gas for 30 minutes to make sure there is no oxygen in the reactor atmosphere. 0.0032 M aqueous CuCl₂ solution of 360 mL is added to the reactor as precursor. 0.35 M aqueous NaOH solution of 12 mL is added to this mixture and 0.1 M sodium ascorbate of 12 mL is added to the reactor, after which the solution turns yellowish-orange in color, it is allowed to stir for 45 minutes. The synthesis is stopped and washed using ethanol three times (sonicate and centrifuge) to obtain medium-sized $Cu_2O$ nanocubes (90-120 nm edge lengths).

Smaller $Cu_2O$ nanocubes of $33 \pm 6$ nm edge length were synthesized using chemical reduction method performed at room temperature (~20 °C). An aqueous $CuCl_2$ solution of 120 mL of 0.0032 M is added to a 250 mL three-neck round bottom flask, which is put in an inert environment filled with nitrogen for 45 min. We added 4 mL of 0.35 M aqueous NaOH solution and 0.1 M sodium ascorbate (reducing agent) was then added (4 mL). The solution subsequently became bright yellow, suggesting that $Cu_2O$ nanoparticles are formed giving smaller $Cu_2O$ nanocubes of $33 \pm 6$ nm edge length. After 45 minutes $Cu_2O$ nanocubes were washed using same washing procedure mentioned above and used in photocatalytic experiments.

**Characterizations of $Cu_2O$ nanospheres and nanocubes**

The synthesized $Cu_2O$ spherical and cubical particles were characterized using UV-Vis-near IR extinction spectroscopy, X-ray diffraction (XRD) analysis, transmission electron microscopy (TEM), and scanning electron microscopy (SEM). All UV-Vis-near IR extinction spectra were taken using an Agilent Cary 60 Spectrophotometer. XRD patterns were acquired using a Philips X-Ray diffractometer (Phillips PW 3710 MPD, PW2233/20 X-Ray tube, Copper tube detector – wavelength - 1.5418 Angstroms), operating at 45 KW, 40 mA. The SEM images were taken using an FEI Quanta 600 F. The TEM images were taken using JEOL JEM-2100 TEM and Thermo Fisher Scientific Titan Themis 200 G2 aberration-corrected TEM. The JEOL JEM-2100 system is equipped with a LaB6 gun and an accelerating voltage of 200 kV. The Titan Themis 200 system is equipped with a Schottky field-emission electron gun and operated at 200 kV.

**II. Experimental procedure for performing photocatalytic experiments**

Photocatalytic MB degradation reaction conditions: A 6 mL quartz tube and the Luzchem Exposure Panels provided with green LED lamps (Figure S2) were used as a reaction chamber and the visible-light source, respectively. To measure the photocatalytic activity of $Cu_2O$ nanospheres and nanocubes of different sizes, the reaction mixture containing 1 mM concentration of MB and 5.8 mg of $Cu_2O$ photocatalyst in 4 ml of solvent (dimethylformamide, DMF) was first allowed to equilibrate at room temperature (~20 °C) for 3 hours in the dark. The reaction mixture was then sparged with air for 30 minutes to keep the soluble oxygen ($O_2$) content in the solvent same for all experiments. The equilibrated mixture under the stirring conditions at room temperature was then exposed to the visible light (i.e., 20 number of green LEDs). The intensity of this light source, when measured at the surface of the photoreactor, was 8.868 mW/cm$^2$. To measure the extent of photodegradation of MB by the $Cu_2O$ nanostructures, the MB concentration (C) in the reaction mixture was quantified as a function of reaction time. The MB absorption value at its peak absorption wavelength (i.e., 665 nm) was used to quantify the MB concentration in the reaction mixture. To obtain the apparent rate constant values, we fitted the concentration versus time profile to the zeroth-order, first-order, and second-order kinetics. Among these fittings, the first-order fittings showed the best fit.

More specific details of the photocatalytic experiments are also provided below.

5.8 mg of $Cu_2O$ nanocatalyst (nanospheres or nanocubes) is uniformly dispersed (sonicated for 2 minutes) in 4 mL of reaction solvent (DMF). The mixture is added to a 6 mL quartz test tube (i.e., photoreactor). For experiments with benzoquinone, 43.2 mg (100 mM) of benzoquinone is added to the reaction mixture. 10 mM methylene blue (MB) solution is made in the same reaction solvent (DMF) used for photocatalytic reaction. 150 uL of 10 mM MB solution is added to the reaction mixture in the quartz test tube and allowed to stir at 1150 rpm and equilibrate at room temperature for three hours in the dark environment. The reactor is transferred to the Luzchem LED Panels (arranged with 4 Luzchem Exposure panels), where LED lamps can be attached as shown in Figure S2. Sampling was done as a function of reaction time by taking 100 µL of the reaction mixture and diluted in 4 mL dilution solvent (DMF). Care was taken to make sure the reaction solvent and diluent are the same. These samples were characterized by UV-Vis to obtain absorption of methylene blue and corresponding $C/C_0$ values as a function of time were obtained. Incident light intensities were measured using Intell Pro Instruments Pro, Smart Sensor purchased from Luzchem Research Inc. The detector is placed exactly where the reactor is placed inside the Luzchem reactor (arranged with 4 Luzchem Exposure panels). Using the Smart sensor and AR823 Digital Lux meter (i.e., purchased from Luzchem Research Inc), the corresponding settings based on the wavelength range of the LED light intensity are measured in Lux. The values are converted to Light intensity in $mW/m^2$ by multiplying measured lux with the calibration factors. For the solvent-dependent studies, the photocatalytic experiments were performed using ethanol (EtOK) as well DMF. In the solvent-dependent studies and benzoquinone experiments (e.g., Figure 7a-b), 10 green LEDs and 10 blue LEDs were used as light source.

### III. Transient absorption measurements

Transient absorption measurements: Optical pump-probe experiments were performed using ~100 fs pulses from a 1-kHz laser amplifier, with a pump center wavelength of 400 nm and a probe center wavelength of 800 nm. Photoexcitation by the pump pulse is above the $Cu_2O$ bandgap and created photoexcited charge carriers. The probe pulse determines the subsequent photocarrier dynamics predominantly through free-carrier absorption. All samples are mixed with a ~1% concentration by mass into a KBr matrix and compressed into a semitransparent disc and measured at room temperature. More specific details are also provided below.

Transient absorption measurements were performed using ~100-fs pulses from a laser amplifier with a 1-kHz repetition rate. The amplifier emits pulses with a center wavelength of 800 nm, which are separated into two replica pulses, time-delayed with a mechanical translation stage with a total relative delay time of ~2 ns. The pump pulses are frequency-doubled in a β-barium borate crystal to 400-nm, so they have photon energy well above the bandgap of $Cu_2O$. The probe pulse remains at 800 nm, close to the band edge of the samples. The probe light transmits (and scatters) through the sample. Each transmission sample is a compressed disk, consisting of approximately 1% of the powdered sample in a KBr matrix, produced in a light vacuum for samples with lower scattering. At the samples, the pump beam has a $1/e^2$ diameter of ~0.35 mm and an average beam power of 1.12 mW, whereas the probe beam has a $1/e^2$ diameter of ~0.1 mm and an average beam power of 0.5 mW. Differential transmission data is recorded in a silicon photodetector, feeding a lock-in-amplifier that is referenced to a mechanical chopper placed in the pump beam and synchronized to a sub-harmonic of the laser amplifier.

## IV. Details of finite-difference time-domain (FDTD) simulations

To implement FDTD simulations, we employed the Lumerical FDTD package.[1] The real (n) and imaginary (k) parts of the refractive indexes used in the simulations for $Cu_2O$, $CeO_2$, $\alpha$-$Fe_2O_3$, and $TiO_2$ are shown in Table S1-S4, respectively. The perfectly matched layer (PML) boundary conditions were used for the simulations in all x, y, and z directions. For the simulations of extinction, scattering, and absorption spectra, the respective cross sections as a function of wavelengths were calculated using the total-field/scattered-field (TFSF) formalism. The incident light source used for these simulations was the Gaussian source in the simulated wavelength region. In the simulations of extinction and absorption spectra, the propagation direction of the incident radiation for cubes simulation (e.g., Figure 4d) is perpendicular or parallel to the principal axes. On implementing the simulations of the magnetic and electric field distributions, a plane wave was used as electromagnetic field incidence with propagation in the x-axis direction, and polarization along the y-axis and the z-axis for the electric field and the magnetic field, respectively.

For the calculation of the relative photocatalytic rates of $Cu_2O$ nanospheres and nanocubes of different sizes, we propose here a simple approximation to relate the photocatalytic rates to the absorption cross section of the $Cu_2O$ nanoparticles. This approximation is similar to the approximation proposed by Ingram et al[2]. for understanding plasmonic resonance-enhanced photocatalysis. In our proposed approximation, the photocatalytic rate (r) on a single nanoparticle for given incident light wavelengths is proportional to the overlap between the absorption spectrum and incident light spectrum.

$$r_{NP} \propto \int \sigma_{abs}(\lambda) I_o(\lambda) d\lambda \qquad (1)$$

where $\sigma_{abs}(\lambda)$ is wavelength-dependent absorption cross and $I_o(\lambda)$ is wavelength-dependent incident light intensity.

The photocatalytic rate for a system that has known weight of $Cu_2O$ nanostructure can then be written as

$$r \propto \int N \times \sigma_{abs}(\lambda) I_o(\lambda) d\lambda \qquad (2)$$

where N is a total number of nanoparticles present in the system of a known weight. The total number (N) of nanoparticles present in the system can be written as

$$N = \frac{m}{\rho V} \qquad (3)$$

where m is a total weight of nanoparticles in the system, and $\rho$ and V are density and volume of a single nanoparticle. Assuming constant density for $Cu_2O$ nanospheres and nanocubes of different sizes, equation (2) can be written as

$$r \propto \int \frac{\sigma_{abs}(\lambda)}{V} I_o(\lambda) d\lambda \qquad (4)$$

From equation (4), for a given system of known total weight of $Cu_2O$ nanostructures, the photocatalytic rate is proportional to the overlap between the volume-normalized absorption spectrum and the incident light spectrum. The relative photocatalytic rate of Cu$_2$O nanostructures of two different sizes (say nanoparticles of Size 1 and nanoparticles of Size 2) can be written as the one shown in equation (5) below. Equation (5) also shows that the overlap between the FDTD-simulated volume-normalized absorption spectrum and incident light spectrum can be used as a critical descriptor for rationally predicting the relative photocatalytic rates. This equation used to predict and model the relative photocatalytic rates of Cu$_2$O nanospheres and nanocubes of different sizes shown in Figure 5 (i.e., the volcano plots) of the main draft.

$$\frac{r_1}{r_2} \propto \frac{\int \left(\frac{\sigma_{abs}(\lambda)}{V}\right)_1 I_o(\lambda) d\lambda}{\int \left(\frac{\sigma_{abs}(\lambda)}{V}\right)_2 I_o(\lambda) d\lambda} \quad (5)$$

**Supplementary Information Figures**

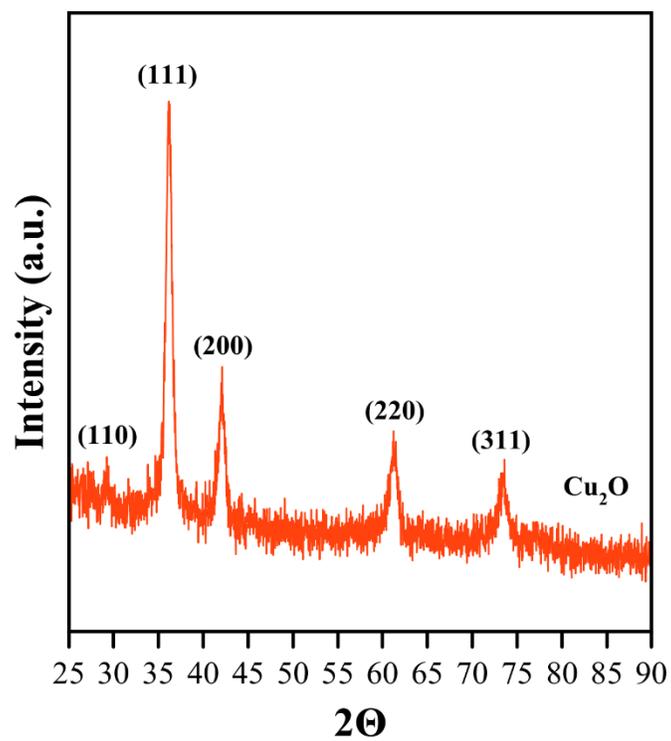

**Figure S1a.** Representative X-ray diffraction pattern of large $Cu_2O$ nanocubes of 286 ± 47 nm edge length synthesized using chemical reduction method.

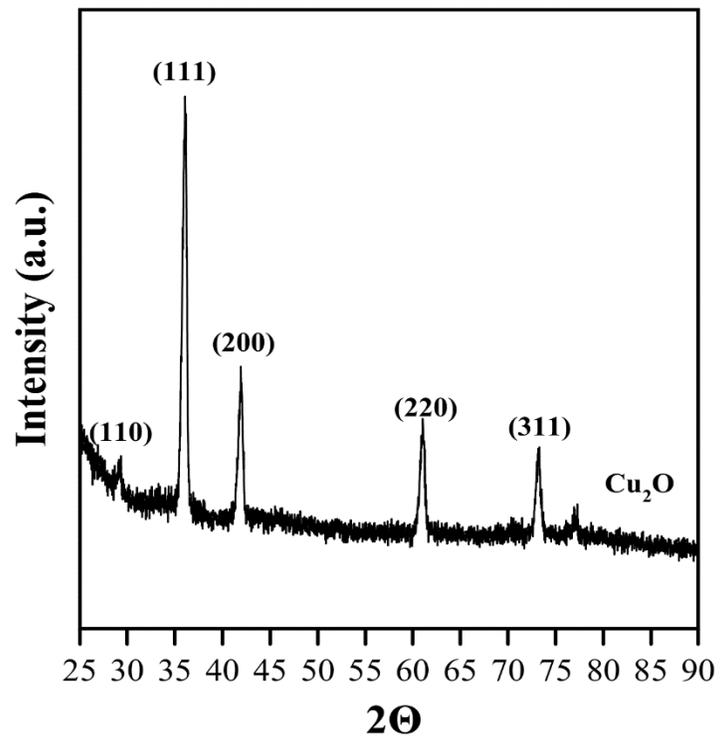

**Figure S1b.** Representative X-ray diffraction pattern of smaller $Cu_2O$ nanospheres of $42 \pm 6$ nm diameter synthesized using microemulsion method.

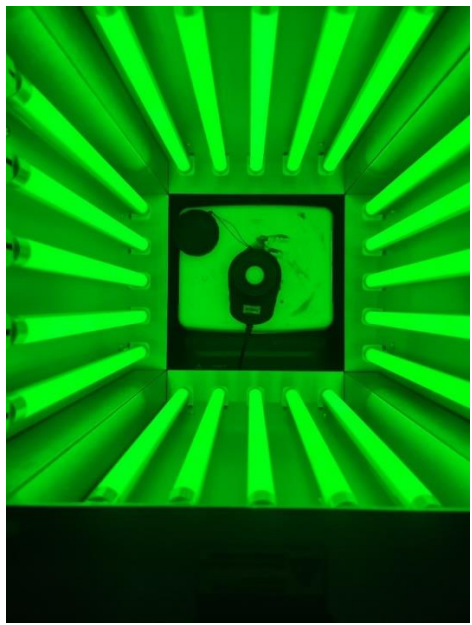

**Figure S2.** Photoreactor reactor set up of Luzchem Exposure Panels connected with green LED lamps.

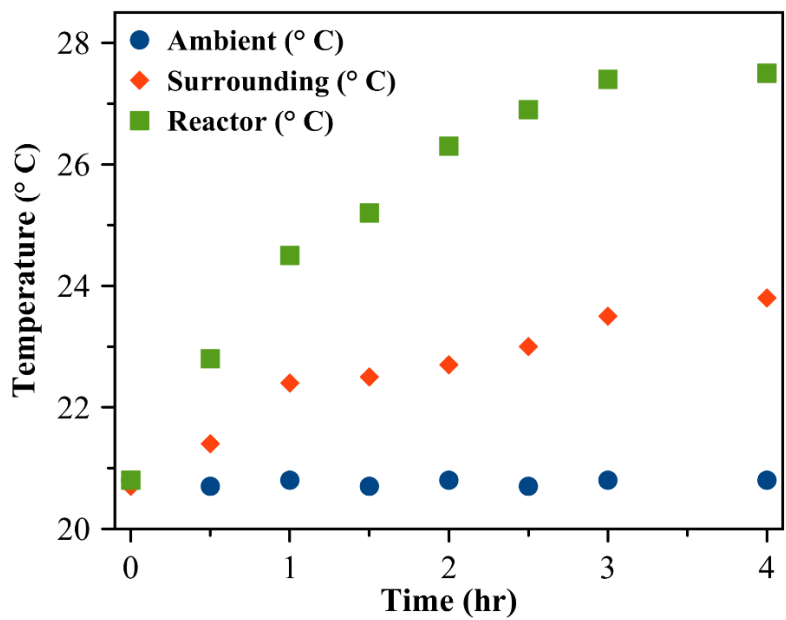

**Figure S3a.** Temperature profile measured as a function of irradiation time during photocatalytic degradation of MB under blank conditions in the absence of photocatalyst. The data represented by green squares, red diamonds, and blue circles show the temperature of the sample in the photoreactor, temperature of the reactor surrounding, and ambient room temperature of the laboratory, respectively.

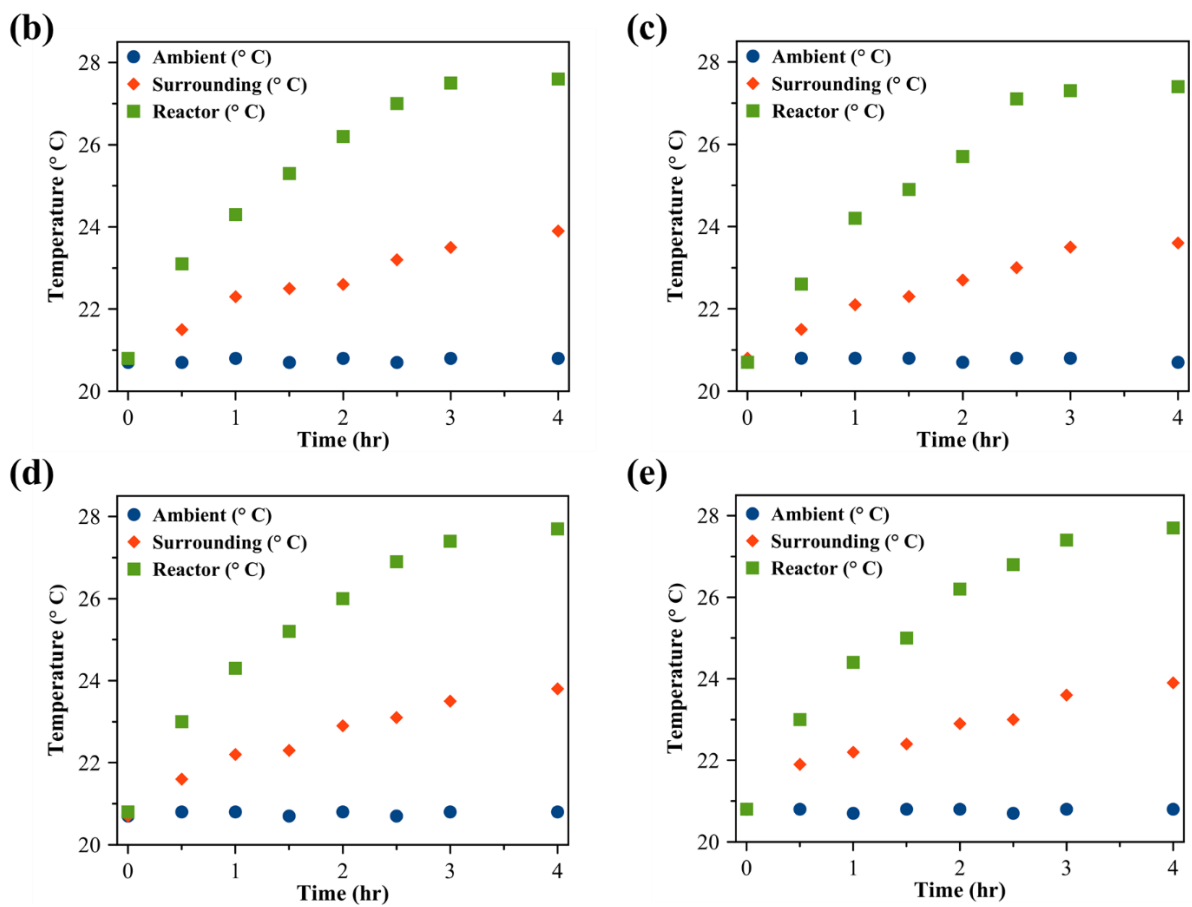

**Figure S3b-e.** Temperature profile measured as a function of irradiation time during photocatalytic degradation of MB for different conditions: (**b**) using $Cu_2O$ nanospheres of $42 \pm 6$ nm diameter, (**c**) $Cu_2O$ nanocubes of $92 \pm 13$ nm edge length. (**d**) $Cu_2O$ nanocubes of $286 \pm 47$ nm edge length. (**e**) using $Cu_2O$ nanospheres of $145 \pm 13$ nm diameter.

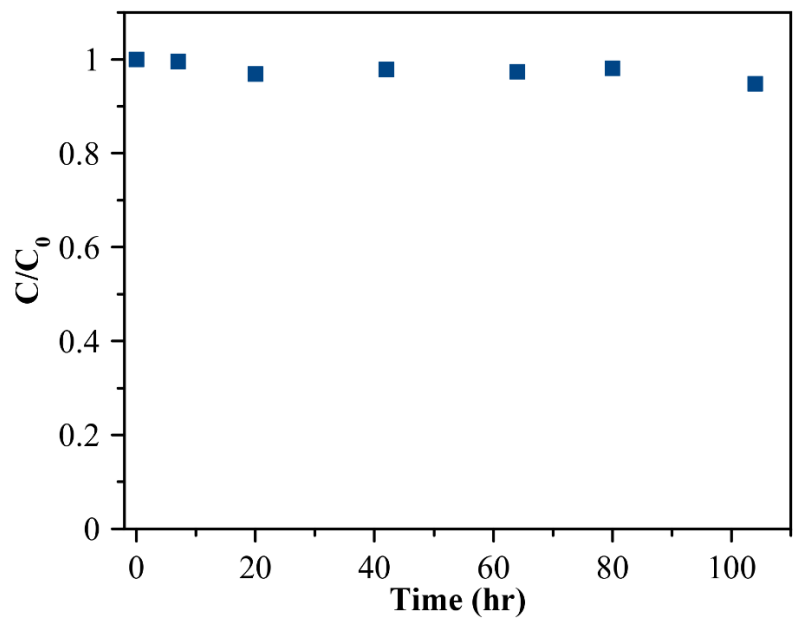

**Figure S3f.** Pot of C/C₀ of methylene blue versus reaction time measured during heating experiments at 60 °C in the presence of large $Cu_2O$ cubes under dark conditions (i.e., in the absence of LED light exposure).

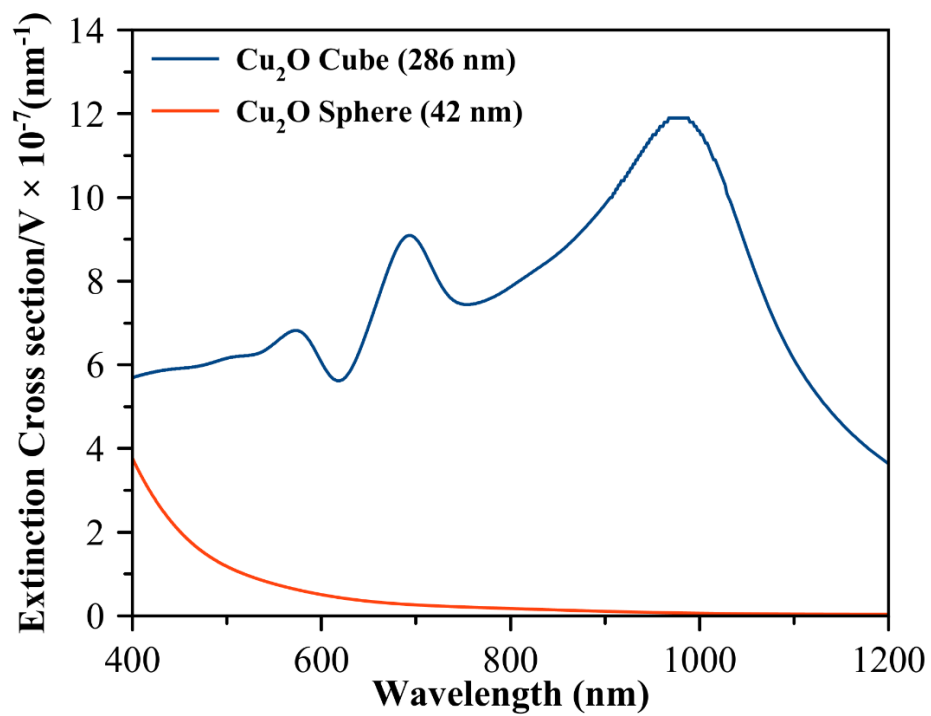

**Figure S4a.** FDTD-simulated volume-normalized extinction cross section of large Cu$_2$O nanocube of 286 nm edge length and small Cu$_2$O nanospheres of 42 nm diameter.

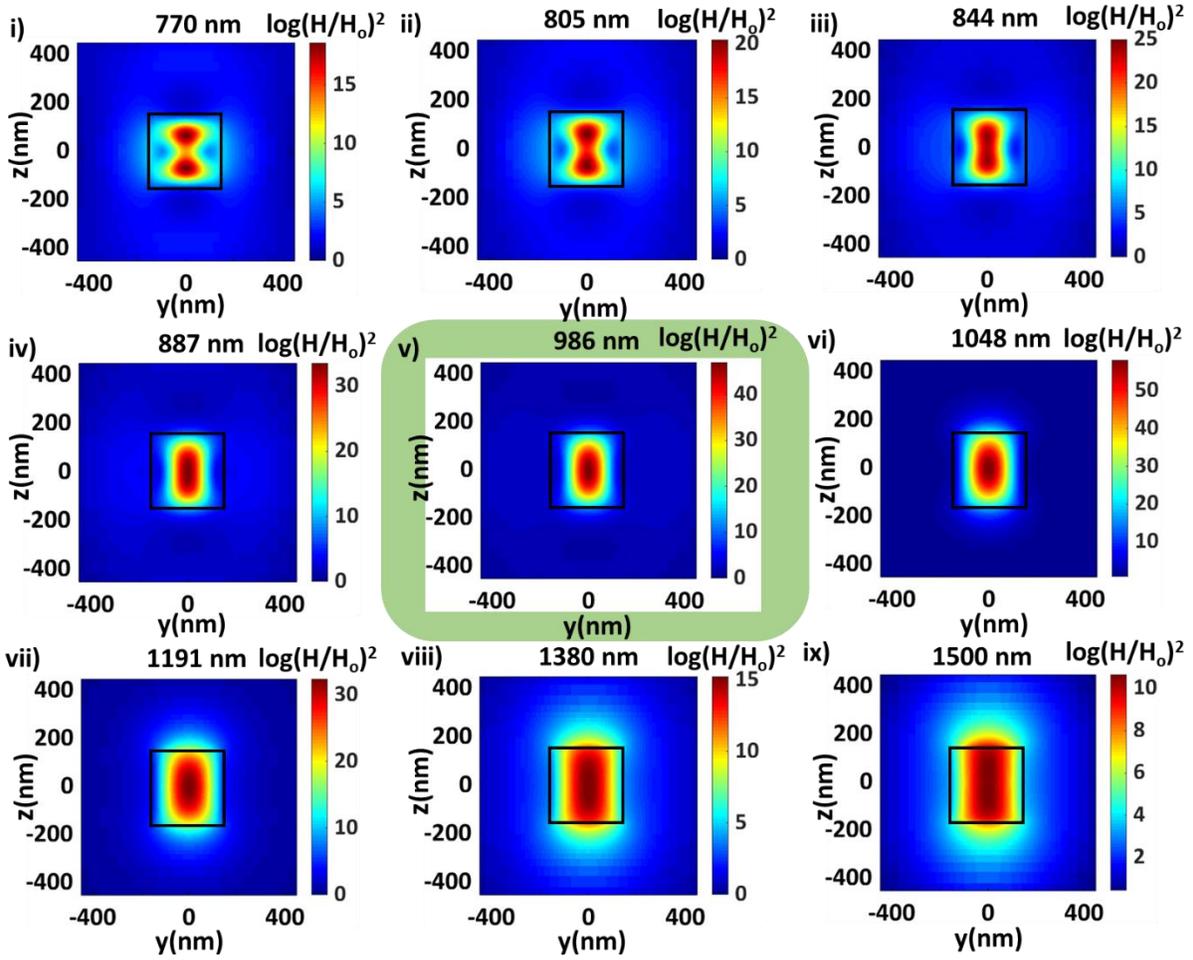

**Figure S4b (I).** FDTD-simulated spatial distribution of enhancement in magnetic field intensity [$H^2/H_0^2$] in YZ plane at different wavelengths across the lowest energy Mie resonance peak wavelength (i.e., 986 nm) for $Cu_2O$ cube of 286 nm edge length.

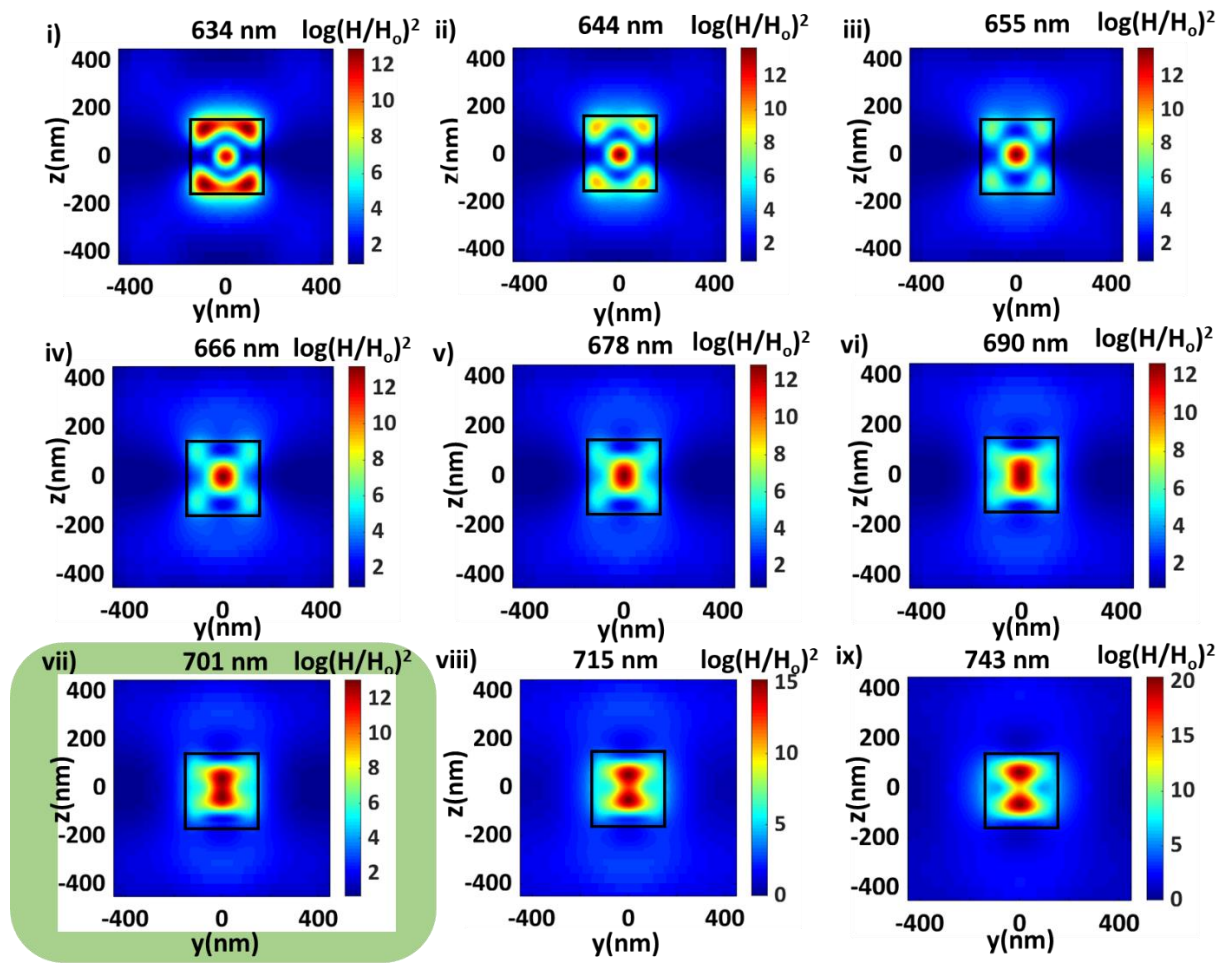

**Figure S4b (II)**. FDTD-simulated spatial distribution of enhancement in magnetic field intensity [$H^2/H_0^2$] in YZ plane at different wavelengths across the second lowest energy Mie resonance peak wavelength (i.e., 701 nm) for $Cu_2O$ cube of 286 nm edge length.

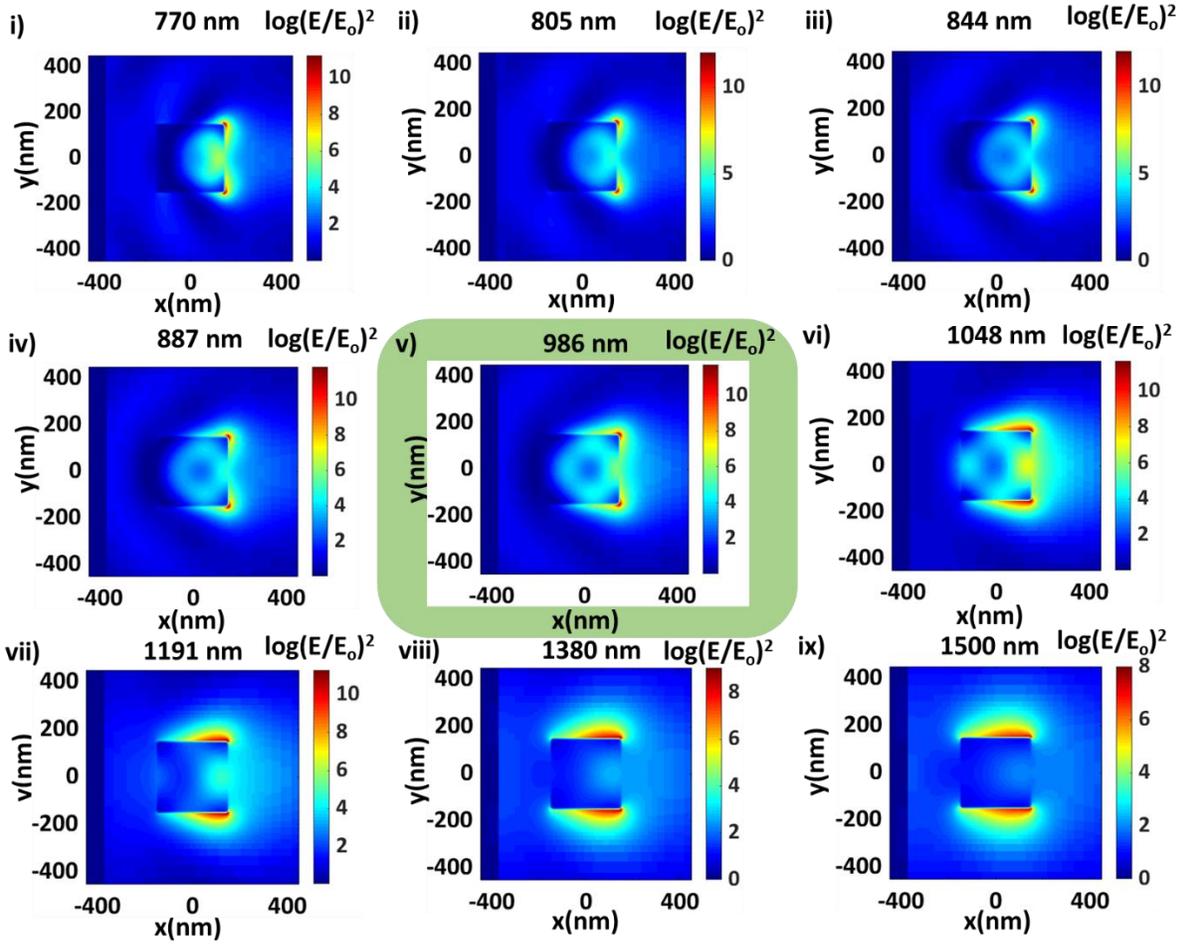

**Figure S4c (I).** FDTD-simulated spatial distribution of enhancement in electric field intensity [$E^2/E_0^2$] in XY plane at different wavelengths across the lowest energy Mie resonance peak wavelength (i.e., 986 nm) for $Cu_2O$ cube of 286 nm edge length.

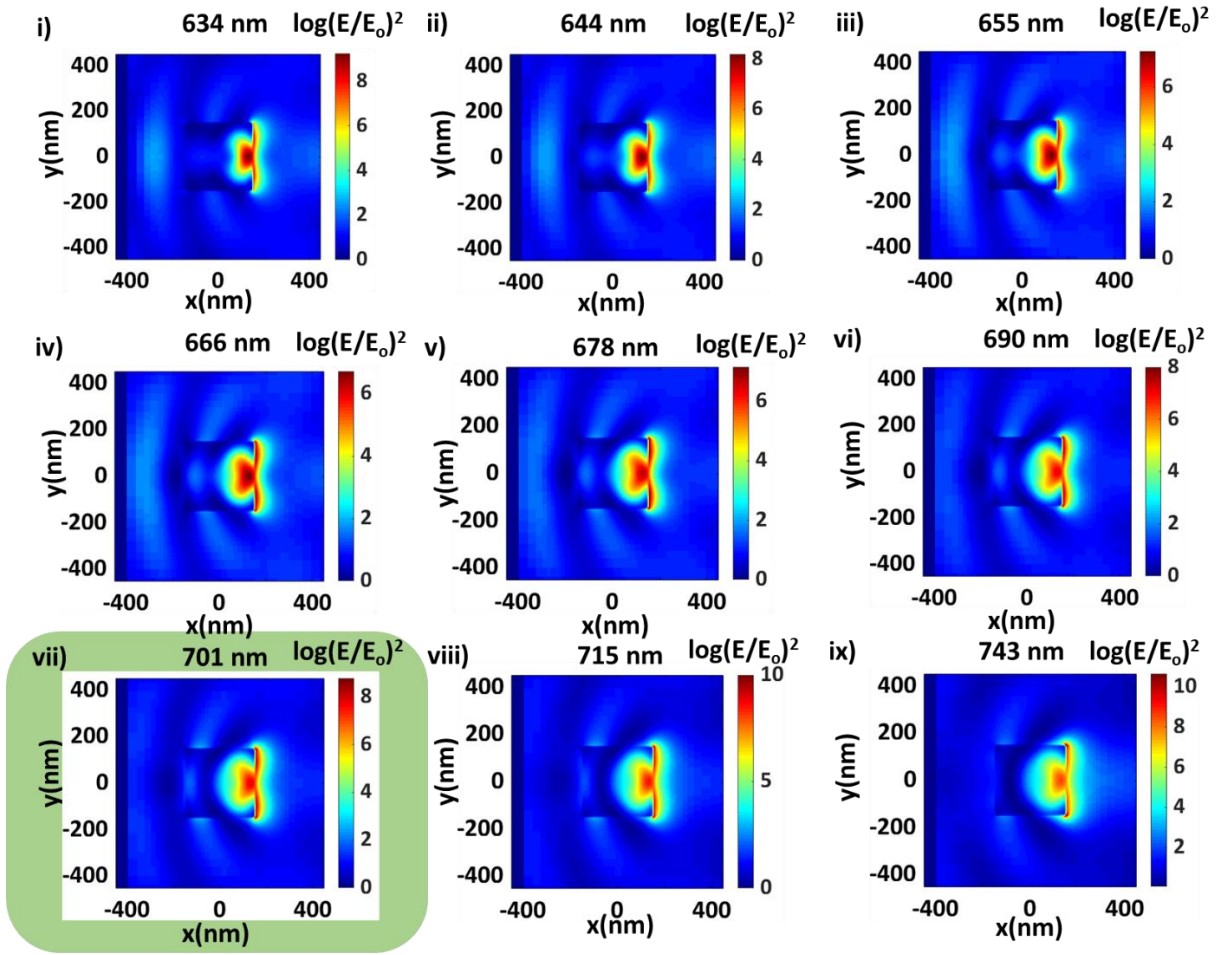

**Figure S4c (II).** FDTD-simulated spatial distribution of enhancement in electric field intensity [$E^2/E_0^2$] in XY plane at different wavelengths across the second lowest energy Mie resonance peak wavelength (i.e., 701 nm) for $Cu_2O$ cube of 286 nm edge length.

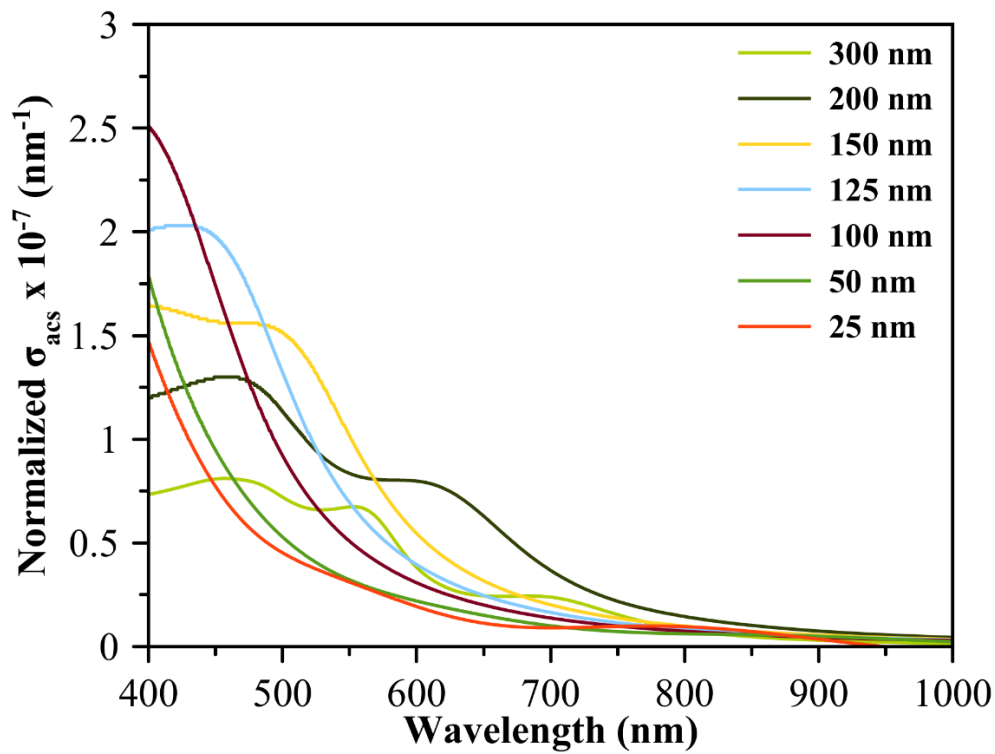

**Figure S5a**. FDTD-simulated volume-normalized absorption cross section ($\sigma_{acs}$) as a function of incident light wavelength for $Cu_2O$ spherical nanoparticles of different sizes.

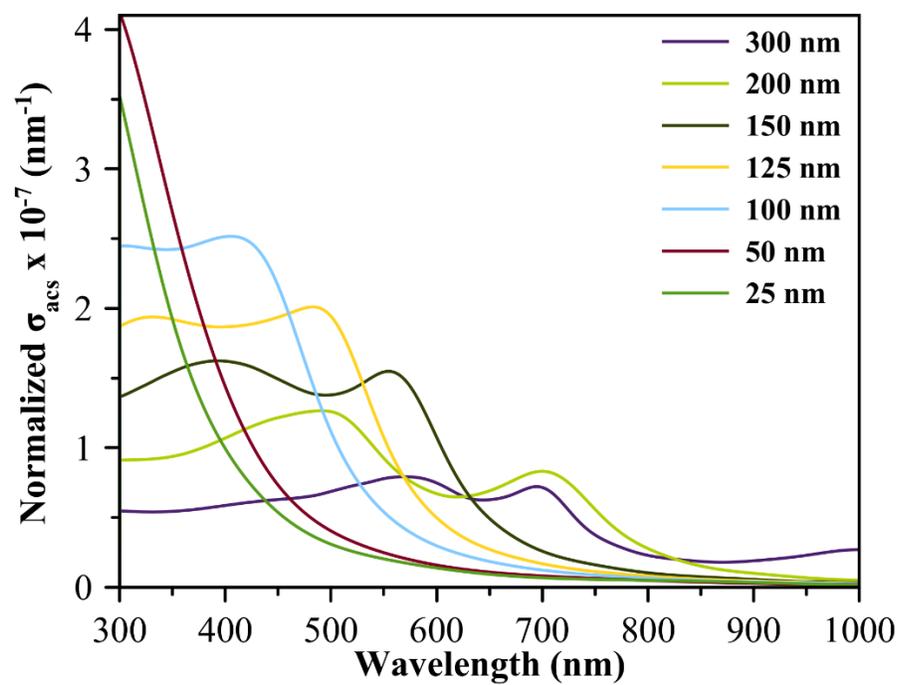

**Figure S5b**. FDTD-simulated volume-normalized absorption cross section ($\sigma_{acs}$) as a function of incident light wavelength for $Cu_2O$ nanocubes of different sizes.

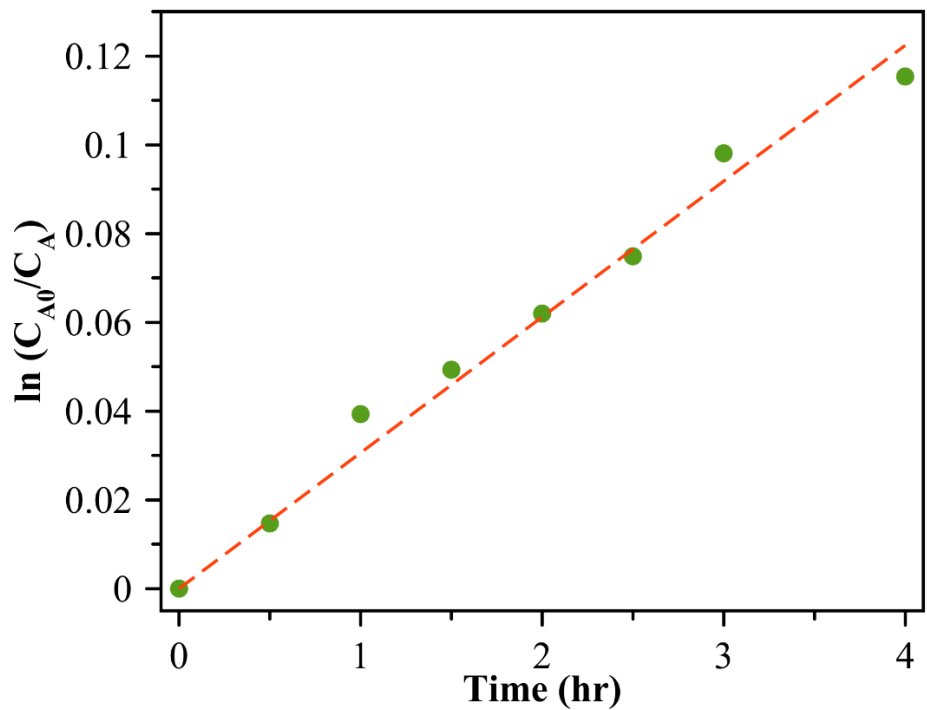

**Figure S5c.** Representative plot of empirical fit with first-order $\ln(C_{A0}/C_A)$ versus irradiation time for photocatalytic degradation of MB under green light illumination using $Cu_2O$ nanospheres of $42 \pm 6$ nm diameter.

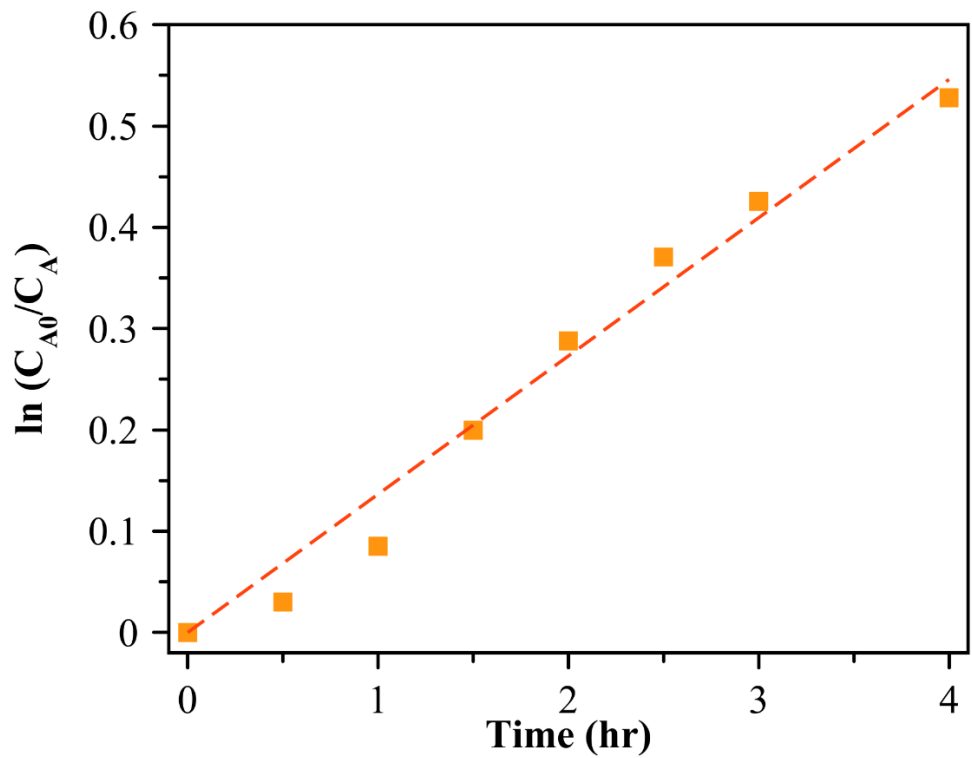

**Figure S5d.** Representative plot of empirical fit with first-order ln($C_{A0}/C_A$) versus irradiation time for photocatalytic degradation of MB under green light illumination using $Cu_2O$ nanocubes of 286 ± 47 nm edge length.

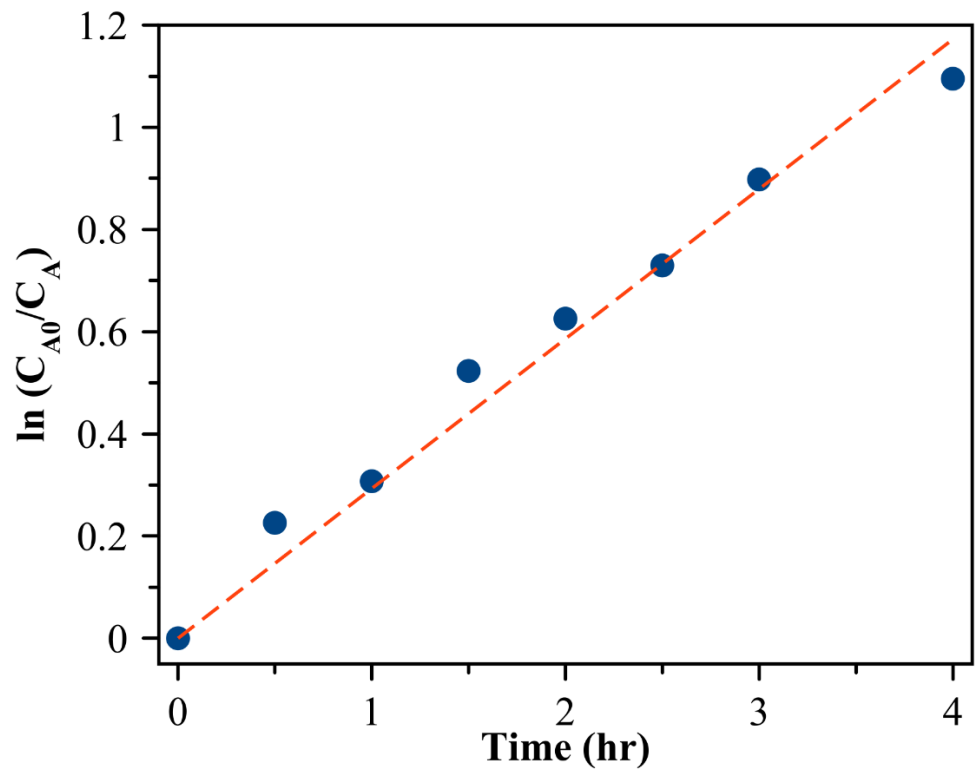

**Figure S5e.** Representative plot of empirical fit with first-order ln($C_{A0}/C_A$) versus irradiation time for photocatalytic degradation of MB under green light illumination using $Cu_2O$ nanospheres of 145 ± 30 nm diameter.

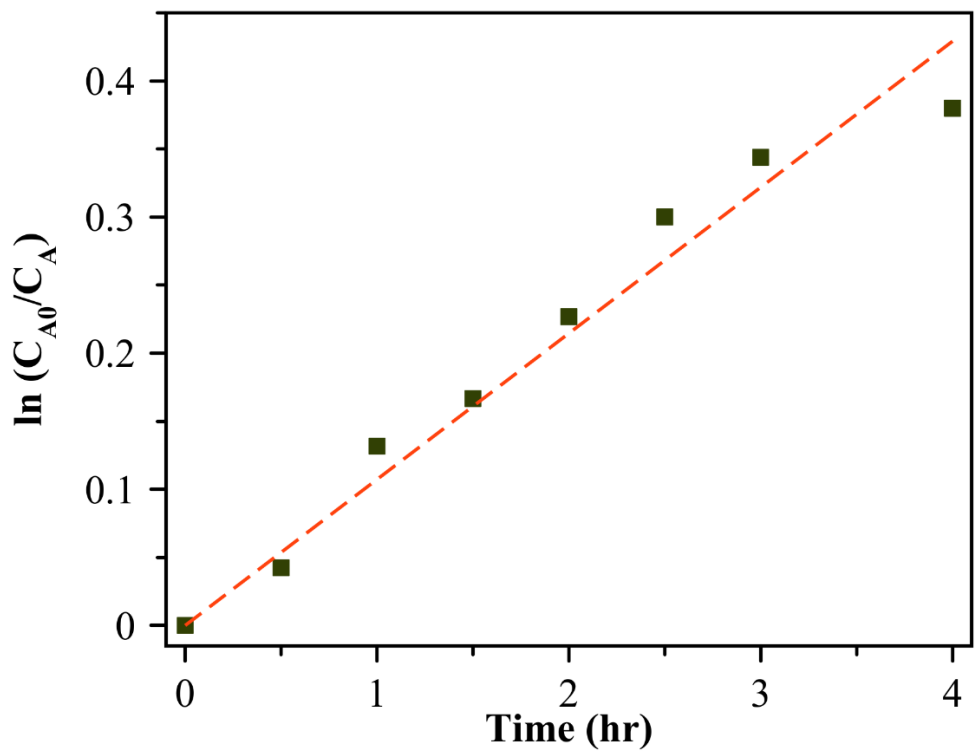

**Figure S5f.** Representative plot of empirical fit with first-order ln($C_{A0}/C_A$) versus irradiation time for photocatalytic degradation of MB under green light illumination using $Cu_2O$ nanocubes of 92 ± 13 nm edge length.

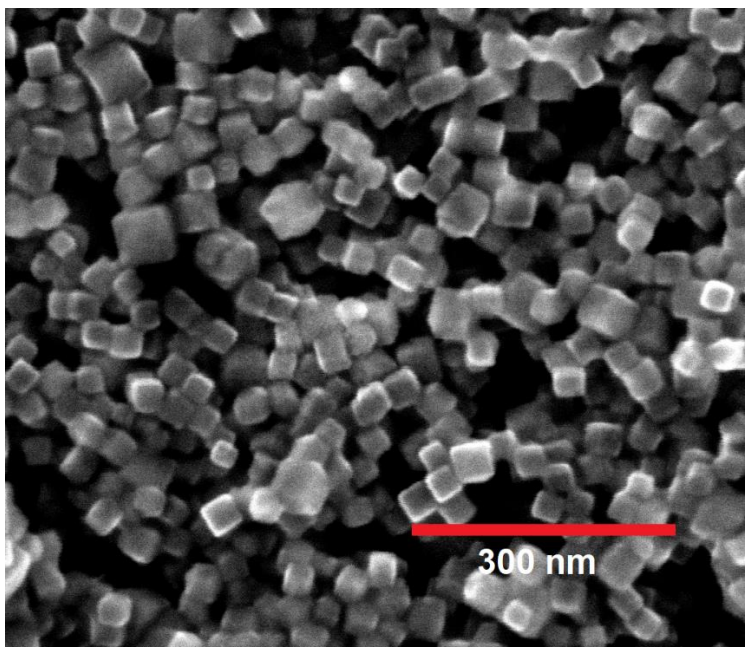

**Figure S6a.** Representative transmission electron microcopy image of Cu$_2$O nanocubes of 33 ± 6 nm edge length.

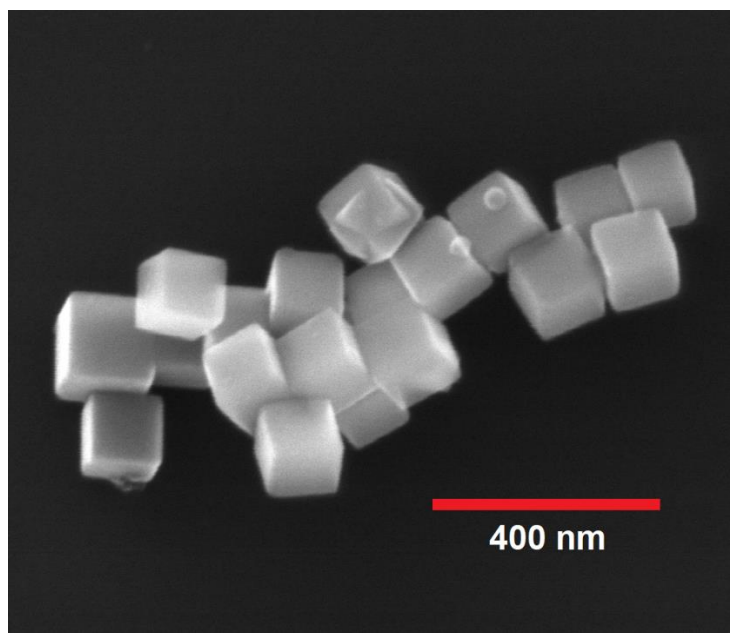

**Figure S6b.** Representative transmission electron microcopy image of Cu$_2$O nanocubes of 118 ± 21 nm edge length.

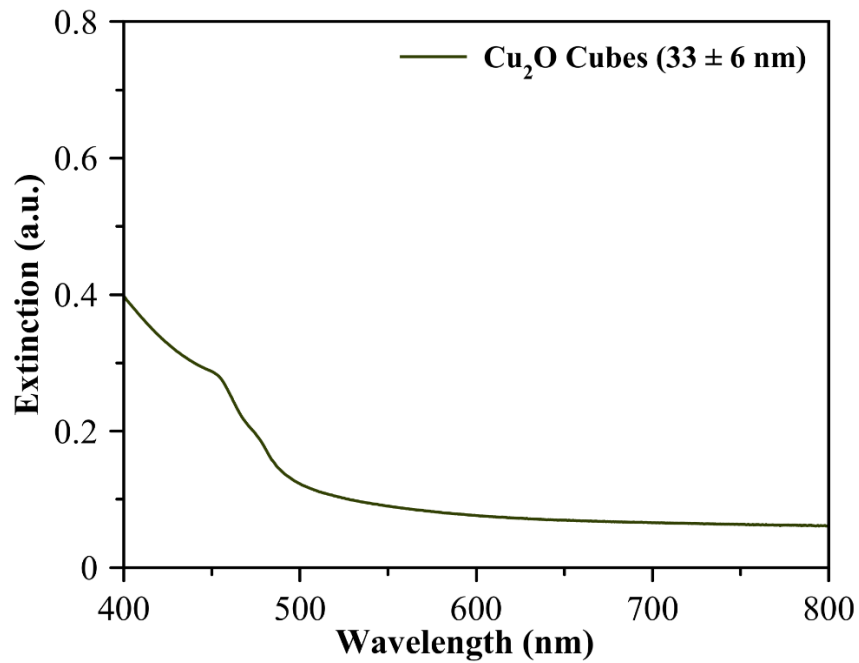

**Figure S6c.** UV-Vis extinction spectra of Cu$_2$O nanocubes of 33 ± 6 nm edge length.

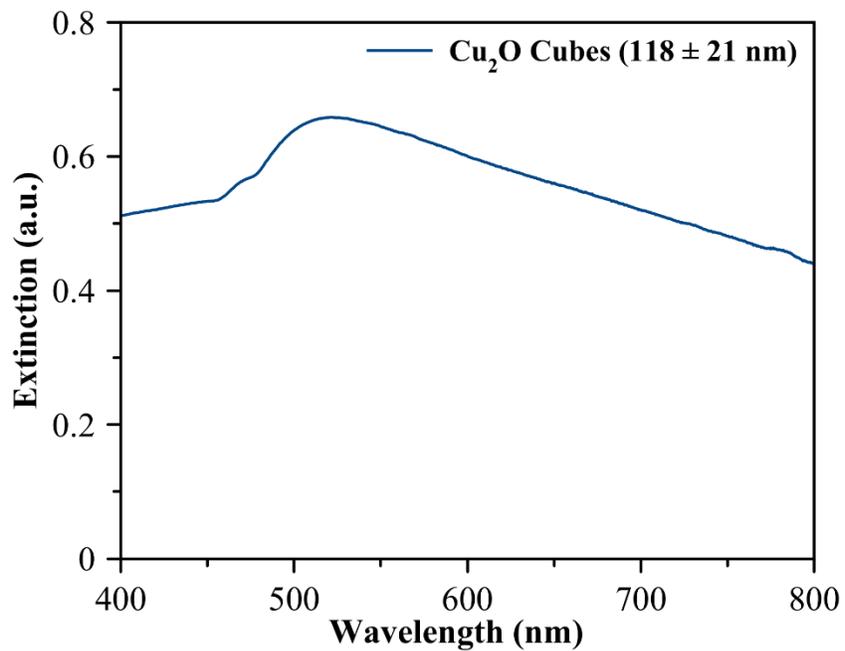

**Figure S6d.** UV-Vis extinction spectra of Cu$_2$O nanocubes of 118 ± 21 nm edge length.

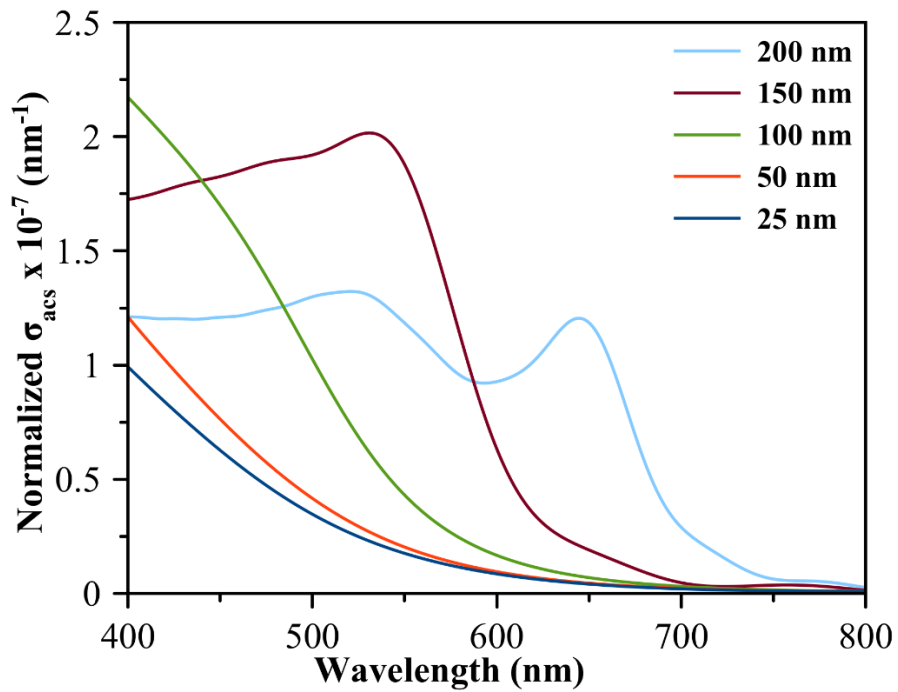

**Figure S7a**. FDTD-simulated volume-normalized absorption cross section ($\sigma_{acs}$) as a function of incident light wavelength for α-$Fe_2O_3$ nanospherical particles of different sizes.

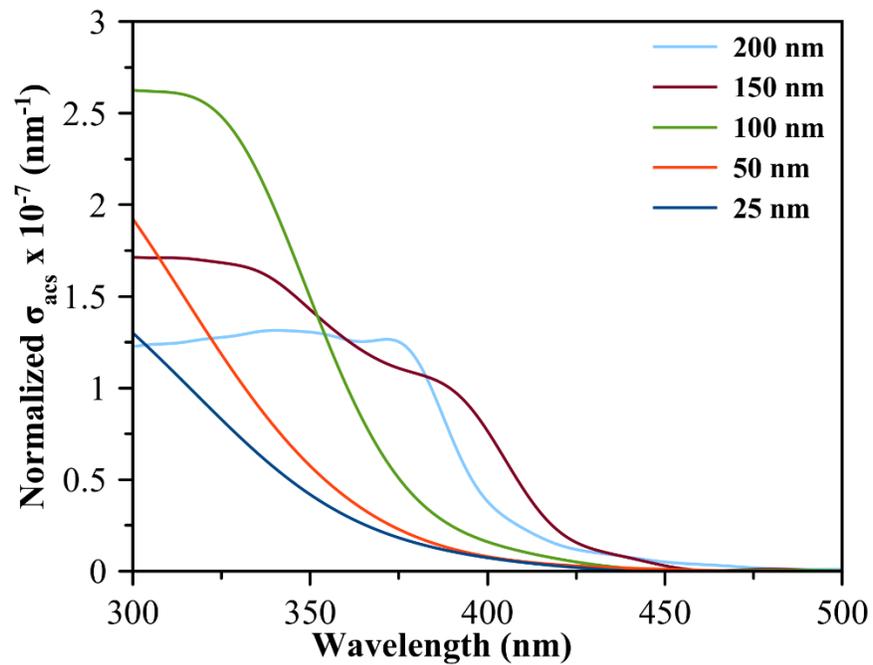

**Figure S7b**. FDTD-simulated volume-normalized absorption cross section ($\sigma_{acs}$) as a function of incident light wavelength for $CeO_2$ spherical nanoparticles of different sizes.

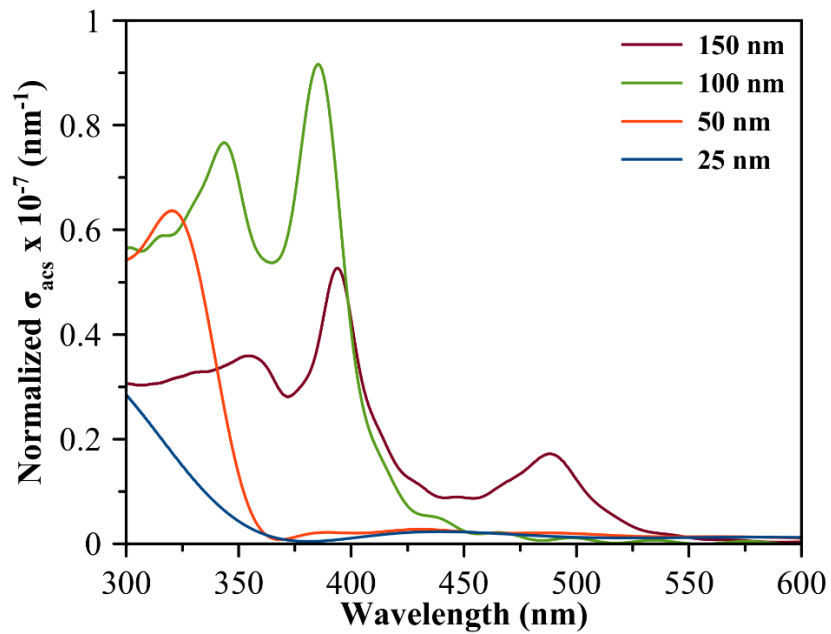

**Figure S7c**. FDTD-simulated volume-normalized absorption cross section ($\sigma_{acs}$) as a function of incident light wavelength for $TiO_2$ spherical nanoparticles of different sizes.

**Table S1**. Fitted first-order rate constant values obtained from the photocatalytic degradation of MB using $Cu_2O$ nanospheres and nanocubes of different sizes.

| $Cu_2O$ nanostructures | Rate Constant, k $(hr^{-1})$ |
|---|---|
| **Spheres (42±6 nm)** | 0.0335 ± 0.0041 |
| **Cubes (92±13 nm)** | 0.1130 ± 0.0030 |
| **Cubes (286±47 nm)** | 0.1408 ± 0.0060 |
| **Spheres (145±13 nm)** | 0.3271 ± 0.0401 |

**Table S2**. The real (n) and imaginary (k) parts of refractive index values of $Cu_2O$ used in the simulations.[3]

|  | $Cu_2O$ | |
| --- | --- | --- |
| λ (nm) | n | k |
| 300 | 2 | 1.85 |
| 350 | 2.4 | 1.44 |
| 400 | 2.8 | 0.99 |
| 450 | 3.06 | 0.6 |
| 500 | 3.12 | 0.35 |
| 550 | 3.1 | 0.19 |
| 600 | 3.02 | 0.13 |
| 650 | 2.9 | 0.1 |
| 700 | 2.83 | 0.083 |
| 750 | 2.77 | 0.07 |
| 800 | 2.7 | 0.06 |
| 850 | 2.66 | 0.053 |
| 900 | 2.63 | 0.048 |
| 950 | 2.61 | 0.043 |
| 1000 | 2.6 | 0.04 |
| 1100 | 2.59 | 0.033 |
| 1200 | 2.58 | 0.027 |
| 1300 | 2.57 | 0.021 |
| 1400 | 2.57 | 0.017 |
| 1500 | 2.57 | 0.013 |
| 2000 | 2.56 | 0.002 |

**Table S3**. The real (n) and imaginary (k) parts of refractive index values of $CeO_2$ used in the simulations.[4]

| | CeO₂ | | | | | | | | | | |
|---|---|---|---|---|---|---|---|---|---|---|---|
| λ (nm) | n | k | λ (nm) | n | k | λ (nm) | n | k | λ (nm) | n | k |
| 401 | 2.47 | 0.11 | 624 | 2.16 | 0.00 | 854 | 2.13 | 0.00 | 929 | 2.12 | 0.00 |
| 410 | 2.41 | 0.08 | 635 | 2.15 | 0.00 | 864 | 2.13 | 0.00 | 940 | 2.12 | 0.00 |
| 421 | 2.38 | 0.03 | 646 | 2.15 | 0.00 | 876 | 2.13 | 0.00 | 951 | 2.12 | 0.00 |
| 429 | 2.36 | 0.02 | 657 | 2.15 | 0.00 | 886 | 2.13 | 0.00 | 962 | 2.11 | 0.00 |
| 439 | 2.31 | 0.01 | 668 | 2.15 | 0.00 | 897 | 2.12 | 0.00 | 973 | 2.11 | 0.00 |
| 450 | 2.29 | 0.01 | 679 | 2.15 | 0.00 | 908 | 2.12 | 0.00 | 984 | 2.11 | 0.00 |
| 461 | 2.28 | 0.00 | 690 | 2.15 | 0.00 | 919 | 2.12 | 0.00 | 996 | 2.11 | 0.00 |
| 471 | 2.25 | 0.00 | 701 | 2.15 | 0.00 | 929 | 2.12 | 0.00 | 929 | 2.12 | 0.00 |
| 483 | 2.24 | 0.00 | 712 | 2.14 | 0.00 | 940 | 2.12 | 0.00 | 940 | 2.12 | 0.00 |
| 494 | 2.22 | 0.00 | 722 | 2.15 | 0.00 | 951 | 2.12 | 0.00 | 951 | 2.12 | 0.00 |
| 504 | 2.21 | 0.00 | 733 | 2.15 | 0.00 | 962 | 2.11 | 0.00 | 962 | 2.11 | 0.00 |
| 516 | 2.20 | 0.00 | 744 | 2.14 | 0.00 | 973 | 2.11 | 0.00 | 973 | 2.11 | 0.00 |
| 526 | 2.19 | 0.00 | 755 | 2.14 | 0.00 | 984 | 2.11 | 0.00 | 984 | 2.11 | 0.00 |
| 537 | 2.18 | 0.00 | 766 | 2.14 | 0.00 | 996 | 2.11 | 0.00 | 996 | 2.11 | 0.00 |
| 548 | 2.18 | 0.00 | 777 | 2.14 | 0.00 | 854 | 2.13 | 0.00 | 929 | 2.12 | 0.00 |
| 559 | 2.18 | 0.00 | 788 | 2.14 | 0.00 | 864 | 2.13 | 0.00 | 940 | 2.12 | 0.00 |
| 570 | 2.18 | 0.00 | 799 | 2.13 | 0.00 | 876 | 2.13 | 0.00 | 951 | 2.12 | 0.00 |
| 581 | 2.17 | 0.00 | 810 | 2.14 | 0.00 | 886 | 2.13 | 0.00 | 962 | 2.11 | 0.00 |
| 592 | 2.16 | 0.00 | 821 | 2.13 | 0.00 | 897 | 2.12 | 0.00 | 973 | 2.11 | 0.00 |
| 602 | 2.16 | 0.00 | 831 | 2.13 | 0.00 | 908 | 2.12 | 0.00 | 984 | 2.11 | 0.00 |
| 613 | 2.16 | 0.00 | 842 | 2.13 | 0.00 | 919 | 2.12 | 0.00 | 996 | 2.11 | 0.00 |

**Table S4**. The real (n) and imaginary (k) parts of refractive index values of α-Fe$_2$O$_3$ used in the simulations.[5]

|         | α-Fe$_2$O$_3$ |       |
| ------- | ------------- | ----- |
| λ (nm)  | n             | k     |
| 400     | 2.756         | 1.294 |
| 450     | 3.181         | 1.02  |
| 500     | 3.282         | 0.675 |
| 550     | 3.318         | 0.498 |
| 600     | 3.265         | 0.149 |
| 650     | 3.074         | 0.057 |
| 700     | 2.972         | 0.031 |
| 750     | 2.903         | 0.021 |
| 800     | 2.853         | 0.02  |
| 850     | 2.824         | 0.027 |
| 900     | 2.805         | 0.024 |
| 950     | 2.789         | 0.022 |
| 1000    | 2.775         | 0.015 |
| 1050    | 2.759         | 0.011 |
| 1100    | 2.745         | 0.011 |
| 1150    | 2.734         | 0.01  |
| 1200    | 2.723         | 0.011 |

**Table S5**. The real (n) and imaginary (k) parts of refractive index values of $TiO_2$ used in the simulations.[6,7]

| | | | | TiO$_2$ | | | | |
|---|---|---|---|---|---|---|---|---|
| λ (nm) | n | k | λ (nm) | n | k | λ (nm) | n | k |
| 180 | 1.37 | 1.998 | 330 | 5.291 | 1.5698 | 480 | 3.08 | 0.0001 |
| 190 | 1.535 | 1.831 | 340 | 4.969 | 1.0926 | 490 | 3.054 | 0.0001 |
| 200 | 1.536 | 1.696 | 350 | 4.477 | 0.6508 | 500 | 3.03 | 0.0001 |
| 210 | 1.46 | 1.65 | 360 | 3.87 | 0.251 | 510 | 3.014 | 0.0001 |
| 220 | 1.433 | 1.806 | 370 | 3.661 | 0.033 | 520 | 3 | 0.0001 |
| 230 | 1.443 | 2.084 | 380 | 3.498 | 0.0001 | 530 | 2.985 | 0.0001 |
| 240 | 1.363 | 2.454 | 390 | 3.375 | 0.0001 | 540 | 2.97 | 0.0001 |
| 250 | 1.365 | 2.847 | 400 | 3.286 | 0.0001 | 550 | 2.954 | 0.0001 |
| 260 | 1.627 | 3.197 | 410 | 3.225 | 0.0001 | 560 | 2.94 | 0.0001 |
| 270 | 1.952 | 3.432 | 420 | 3.186 | 0.0001 | 570 | 2.929 | 0.0001 |
| 280 | 3.355 | 3.561 | 430 | 3.162 | 0.0001 | 580 | 2.92 | 0.0001 |
| 290 | 3.835 | 3.535 | 440 | 3.149 | 0.0001 | 590 | 2.91 | 0.0001 |
| 300 | 4.732 | 3.28 | 450 | 3.141 | 0.0001 | 600 | 2.9 | 0.0001 |
| 310 | 5.235 | 2.734 | 460 | 3.13 | 0.0001 | 610 | 2.889 | 0.0001 |
| 320 | 5.391 | 2.076 | 470 | 3.104 | 0.0001 | 620 | 2.88 | 0.0001 |

| TiO$_2$ | | | | | | | | |
|---|---|---|---|---|---|---|---|---|
| λ (nm) | n | k | λ (nm) | n | k | λ (nm) | n | k |
| 630 | 2.875 | 0.0001 | 780 | 2.8 | 0.0001 | 930 | 2.759 | 0.0001 |
| 640 | 2.87 | 0.0001 | 790 | 2.794 | 0.0001 | 940 | 2.76 | 0.0001 |
| 650 | 2.86 | 0.0001 | 800 | 2.79 | 0.0001 | 950 | 2.761 | 0.0001 |
| 660 | 2.85 | 0.0001 | 810 | 2.79 | 0.0001 | 960 | 2.76 | 0.0001 |
| 670 | 2.844 | 0.0001 | 820 | 2.79 | 0.0001 | 970 | 2.755 | 0.0001 |
| 680 | 2.84 | 0.0001 | 830 | 2.785 | 0.0001 | 980 | 2.75 | 0.0001 |
| 690 | 2.835 | 0.0001 | 840 | 2.78 | 0.0001 | 990 | 2.749 | 0.0001 |
| 700 | 2.83 | 0.0001 | 850 | 2.78 | 0.0001 | 1000 | 2.75 | 0.0001 |
| 710 | 2.825 | 0.0001 | 860 | 2.78 | 0.0001 | 1010 | 2.75 | 0.0001 |
| 720 | 2.82 | 0.0001 | 870 | 2.775 | 0.0001 | 1020 | 2.749 | 0.0001 |
| 730 | 2.814 | 0.0001 | 880 | 2.77 | 0.0001 | 1030 | 2.749 | 0.0001 |
| 740 | 2.81 | 0.0001 | 890 | 2.77 | 0.0001 | 1040 | 2.748 | 0.0001 |
| 750 | 2.81 | 0.0001 | 900 | 2.77 | 0.0001 | 1050 | 2.747 | 0.0001 |
| 760 | 2.81 | 0.0001 | 910 | 2.765 | 0.0001 | 1060 | 2.747 | 0.0001 |
| 770 | 2.806 | 0.0001 | 920 | 2.76 | 0.0001 | 1070 | 2.746 | 0.0001 |

| TiO₂ |||||||||
|---|---|---|---|---|---|---|---|---|
| λ (nm) | n | k | λ (nm) | n | k | λ (nm) | n | k |
| 1080 | 2.745 | 0.0001 | 1230 | 2.729 | 0.0001 | 1380 | 2.721 | 0.0001 |
| 1090 | 2.744 | 0.0001 | 1240 | 2.729 | 0.0001 | 1390 | 2.721 | 0.0001 |
| 1100 | 2.742 | 0.0001 | 1250 | 2.728 | 0.0001 | 1400 | 2.72 | 0.0001 |
| 1110 | 2.741 | 0.0001 | 1260 | 2.728 | 0.0001 | 1410 | 2.719 | 0.0001 |
| 1120 | 2.74 | 0.0001 | 1270 | 2.727 | 0.0001 | 1420 | 2.719 | 0.0001 |
| 1130 | 2.739 | 0.0001 | 1280 | 2.727 | 0.0001 | 1430 | 2.718 | 0.0001 |
| 1140 | 2.738 | 0.0001 | 1290 | 2.726 | 0.0001 | 1440 | 2.717 | 0.0001 |
| 1150 | 2.737 | 0.0001 | 1300 | 2.726 | 0.0001 | 1450 | 2.716 | 0.0001 |
| 1160 | 2.736 | 0.0001 | 1310 | 2.725 | 0.0001 | 1460 | 2.715 | 0.0001 |
| 1170 | 2.735 | 0.0001 | 1320 | 2.725 | 0.0001 | 1470 | 2.714 | 0.0001 |
| 1180 | 2.734 | 0.0001 | 1330 | 2.724 | 0.0001 | 1480 | 2.713 | 0.0001 |
| 1190 | 2.733 | 0.0001 | 1340 | 2.724 | 0.0001 | 1490 | 2.711 | 0.0001 |
| 1200 | 2.732 | 0.0001 | 1350 | 2.723 | 0.0001 | 1500 | 2.71 | 0 |
| 1210 | 2.731 | 0.0001 | 1360 | 2.723 | 0.0001 | | | |
| 1220 | 2.73 | 0.0001 | 1370 | 2.722 | 0.0001 | | | |

**References Cited in Supporting Information**